 \documentclass[%
  aip,jcp,
 amsmath,amssymb,
  reprint,%
 ]{revtex4-1}

\usepackage{graphicx}
\usepackage{dcolumn}
\usepackage{bm}

\usepackage{mathtools}
\usepackage{multirow}
\usepackage{makecell}
\usepackage{xcolor}
\usepackage[export]{adjustbox}

\begin{document}

\preprint{APS/123-QED}

\title{Phase behavior and crystal nucleation of hard triangular prisms}

\author{Marjolein de Jager}
\affiliation{Soft Condensed Matter and Biophysics, Debye Institute for Nanomaterials Science, Utrecht University, Utrecht, Netherlands}
\author{Nena Slaats}
\affiliation{Soft Condensed Matter and Biophysics, Debye Institute for Nanomaterials Science, Utrecht University, Utrecht, Netherlands}
\author{Laura Filion}
\affiliation{Soft Condensed Matter and Biophysics, Debye Institute for Nanomaterials Science, Utrecht University, Utrecht, Netherlands}


\date{\today}

\begin{abstract}
The interplay between densification and positional ordering during the process of crystal nucleation is a greatly investigated topic. Even for the simplest colloidal model -- hard spheres -- there has been much debate regarding the potential foreshadowing of nucleation by significant fluctuations in either local density or local structure. 
Considering anisotropic particles instead of spheres adds a third degree of freedom to the self-organization process of crystal nucleation: orientational ordering.
Here, we investigate the crystal nucleation of hard triangular prisms. 
Using Monte Carlo simulations, we first carefully determine the crystal--fluid coexistence values and calculate the nucleation barriers for two degrees of supersaturation. 
Next, we use brute force simulations to obtain a large set of spontaneous nucleation events.
By studying the time evolution of the local density, positional ordering, and orientational ordering in the region in which the nucleus first arises, we demonstrate that all local order parameters increase simultaneously from the very start of the nucleation process. 
We thus conclude that we observe no precursor for the crystal nucleation of hard triangular prisms.
\end{abstract}

\maketitle


\newcommand{\comment}[1]{{\color{red}{\bf #1}}}

\newcommand{\figwidthA}{0.9\linewidth} 
\newcommand{\figwidthB}{0.6\linewidth}
\newcommand{\figwidthZ}{0.361\linewidth}


\section{Introduction}

Colloidal particles have the ability to self-organize into complex mesoscale structures. An intriguing example of this is the phenomenon of crystal nucleation in which an ordered (crystalline) structure emerges from a disordered fluid phase. This phenomenon is not only important in the field of colloidal science, but also plays a major role in fields concerned with e.g. protein crystallization, ice formation, and polymorph selection in pharmaceuticals. 
Yet, despite its widespread significance, the intricate mechanisms underlying the formation of a crystal nucleus still remain to be fully uncovered.  

In the simplest approach for nucleation, the work associated with nucleation is viewed as the competition between the free-energy gain of transforming into the more stable crystal phase and the free-energy cost of creating a surface. 
However, nucleation is not always as straightforward as the classical pathway that classical nucleation theory (CNT) describes.
Already in 1897 Ostwald introduced his famous two-step scenario for nucleation \cite{ostwald1897studien}, and simple models like Lennard-Jones have been postulated to follow this rule \cite{tenwolde1995numerical}. 
Recently, more exotic models like zeolites have also been shown to nucleate in a two-step fashion \cite{kumar2018two}.
Even for simple hard spheres, there has been extensive discussion on the specifics of the nucleation pathway. 
For example, it has been suggested, e.g. Ref. \onlinecite{russo2012selection},  that it is a two-step process in which fluctuations in positional ordering trigger crystal nucleation.
However, most recent studies demonstrated that the crystal nucleation of hard spheres is indeed a one-step pathway in which densification and positional ordering arise simultaneously \cite{berryman2016early,dejager2023search}.

One of the key characteristics of colloidal particles is their capacity to be synthesized into a wide array of shapes \cite{glotzer2007anisotropy,hueckel2021total}, ranging from spheres to rods to faceted polyhedra. 
The anisotropy of particles adds a third degree of freedom to the self-organization of the particles: orientational ordering. 
It is thus intriguing to study the interplay between densification, positional ordering, and orientational ordering during the crystal nucleation of anisotropic particles.
For example, systems of hard colloidal cubes have been shown to follow a non-classical nucleation pathway in which the spontaneous formation of domains with increased orientational ordering foreshadows the formation of the crystal \cite{sharma2018disorder}. 
These domains of increased cubatic order are facilitated by the high affinity for facet alignment in these systems, which severely lowers the interfacial free energy. 
At phase coexistence conditions, the interfacial free energy of hard cubes is only one-fifth of that for hard spheres \cite{sharma2021low}.
Other, more complex, hard polyhedra, like pentagonal and triangular bipyramids, have also been shown to nucleate in a two-step fashion by first forming a high-density precursor with more pronounced prenucleation order \cite{lee2019entropic}.

In this work, we investigate the crystal nucleation of one of the simplest polyhedra: the triangular prism. Not only is this particle shape experimentally realizable, see e.g. Refs. \onlinecite{dendukuri2006continuous,malikova2002layer,xue2007plasmon,millstone2009colloidal,jin2003controlling,sun2003triangular,xue2008mechanistic}, simulations of hard triangular prisms have revealed that they exhibit a first-order phase transition between isotropic fluid and a honeycomb crystal for some aspect ratios \cite{agarwal2011mesophase}. Hence, using Monte Carlo (MC) simulations of hard triangular prisms, we first carefully determine the phase boundaries of this first-order phase transition. 
Next, we calculate the crystal nucleation barrier for two degrees of supersaturation. Lastly, we obtain numerous spontaneous nucleation events for both degrees of supersaturation via brute force simulations and investigate the onset of crystal nucleation. Similar to Refs. \onlinecite{berryman2016early,dejager2023search}, we focus on the regions in which the nuclei are born and track the time evolution of the local density, positional ordering, and orientational ordering in these regions. We find that all local order parameters increase simultaneously as soon as nucleation starts, indicating that crystal nucleation of hard triangular prisms is a truly spontaneous event with no precursor predicting it.


\section{Model}
\begin{figure}[t!]
     \centering
     \includegraphics[width=0.95\linewidth]{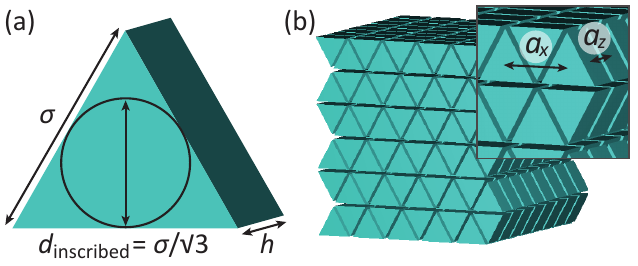}
     \caption{\label{fig:prism}
     Schematic representation of \textbf{(a)} the triangular prism with $h=\sigma/\sqrt{3}$ and \textbf{(b)} its honeycomb crystal with lattice constants $a_x$ and $a_z$. 
     }
\end{figure}

In this work, we consider hard anisotropic particles in the shape of triangular prisms. The two triangular faces of these prisms are equilateral triangles with edge length $\sigma$. The height of each prism is set equal to the diameter of the inscribed circle of the triangular face, i.e. $h=\sigma/\sqrt{3}$ (see Fig. \ref{fig:prism}).
Agarwal and Escobedo determined the phase behavior of these particles in Ref. \onlinecite{agarwal2011mesophase} and found a first-order phase transition between the isotropic fluid phase and a honeycomb crystal. 
Note that for more thin prisms ($h=\sigma/(2\sqrt{3})$) they found a stable smectic phase as an intermediate between the isotropic fluid phase and honeycomb crystal, and for more elongated prisms ($h=2\sigma/\sqrt{3}$) they found an intermediate columnar phase.

\section{Methods}


\subsection{Phase behavior}
To determine the equation of state, we use MC simulations of $N=1344$ particles in the $NPT$ ensemble. In these simulations, we perform isotropic volume changes to determine the equation of state of the fluid phase. For the crystal phase, however, we perform decoupled volume changes, i.e. where the volume changes are performed by independently changing the box size in the $xy$-direction and $z$-direction, such that we can determine the equilibrium height ratio $a_z/a_x$ of the honeycomb crystal (see Fig. \ref{fig:prism}b).

In order to find the coexistence densities of the isotropic fluid and honeycomb crystal, we use thermodynamic integration in combination with the common tangent method. 
For the fluid phase, we use the ideal gas as a reference system and integrate the equation of state to obtain the free energy as a function of the density\cite{frenkelbook}, i.e.
\begin{equation}\label{eq:ffluid}
    \frac{\beta F}{N} = \log\left(\rho\Lambda^3\right) - 1 + \int_0^\rho \frac{\beta P(\rho')-\rho'}{\rho'\,^2} d\rho', 
\end{equation}
where $\Lambda$ is the De Broglie wavelength which we set equal to $\sigma$.
For the crystal phase, we integrate the equation of state from a certain density for which we know the free energy. We obtain this reference free energy using Einstein integration \cite{frenkel1984new} of a perfect honeycomb crystal and correct for finite-size effects \cite{polson2000finite} by considering system sizes of 120, 224, 480, 720, 1344, 2240, and 4080 particles.
In this approach, the free energy of the honeycomb crystal is determined via a thermodynamic integration between the original system and a non-interacting Einstein crystal as a reference system.
This involves performing a series of MC simulations in the $NVT$ ensemble with an effective Hamiltonian given by
\begin{equation}\label{eq:Hein}
    H(\lambda) = U + \lambda U_\text{Ein},
\end{equation}
where $U$ and $U_\text{Ein}$ are the total potential energies of, respectively, the original crystal of hard triangular prisms and the non-interacting Einstein crystal, and $\lambda$ is a parameter that tunes between the original crystal of hard triangular prisms ($\lambda=0$) and the interacting Einstein crystal ($\lambda=1$).
Since we consider anisotropic particles, the potential energy of the non-interacting Einstein crystal consists of two contributions\cite{frenkel1985hard,noya2007phase,smallenburg2012vacancy}: $U_\text{Ein}=U_\text{Ein}^\text{pos}+U_\text{Ein}^\text{rot}$. The first contribution comes from the springs binding the particles to their lattice sites and is given by
\begin{equation}\label{eq:Einpos}
    U_\text{Ein}^\text{pos} = \frac{\alpha}{2\sigma^2} \sum_{i=1}^N ( \mathbf{r}_i-\mathbf{R}_0^{(i)})^2 ,
\end{equation}
where $\alpha$ is the spring constant and $\mathbf{R}_0^{(i)}$ are the lattice positions in the ideal crystal.
In order to prevent any drift of the crystal as a whole, we keep the center of mass of the system fixed during the MC simulations and prevent lattice site hopping by restricting the particles to their Wigner--Seitz cells.
The second contribution comes from the springs aligning the orientations of the particles to their orientation in the ideal crystal
\begin{equation}\label{eq:Einrot}
    U_\text{Ein}^\text{rot} = \frac{\alpha}{2} \sum_{i=1}^N \left[ 2 - (\mathbf{\hat{u}}_z^{(i)}\cdot\mathbf{\hat{z}})^2 - \max_k\left\{(\mathbf{\hat{u}}_k^{(i)}\cdot\mathbf{\hat{y}})^2\right\} \right],
\end{equation}
where $\alpha$ is the same spring constant as in Eq. \eqref{eq:Einpos}, $\mathbf{\hat{u}}_z^{(i)}$ is the normal of the triangular face of particle $i$ and, $\mathbf{\hat{u}}_k^{(i)}$ with $k\in[1,2,3]$ are the 3 normals of the side faces (also referred to as the lateral faces). The second term of the sum penalizes the misalignment of the long axis of the triangular prism and the last term penalizes the misalignment of the lateral faces (see Fig. \ref{fig:prism}b).
Note that, for convenience, we take the square of the product $\mathbf{\hat{u}}_k^{(i)}\cdot\mathbf{\hat{y}}$. This allows us to use the same alignment potential for each particle of the system, even though the honeycomb crystal consists of triangular prisms in two (mirrored) orientations. 
While using this alignment potential could potentially result in incorrectly orientated triangular prisms, in this case the high density of the honeycomb crystal restricts the triangular prisms from performing such a reorientation. However, we note that it will cause a difference of $\log 2$ in the rotational free energy of the non-interacting Einstein crystal.

For a sufficiently high spring constant, the particles are tied very strongly to their ideal lattice positions and orientations such that the system transforms into a non-interacting Einstein crystal at $\lambda=1$. The total free energy of the crystal is then given by 
\begin{equation} 
    \frac{\beta F}{N} = \frac{\beta F_\text{Ein}^\text{rot}}{N}  + 3\log\frac{\Lambda}{d_\alpha} + \frac{1}{N}\log\frac{\rho d_\alpha^3}{N^{3/2}} - \frac{\beta}{N} \int_0^1 \left\langle \frac{\partial H}{\partial \lambda} \right\rangle_\lambda d\lambda , \label{eq:Feintotal}
\end{equation}
where $\langle\dots\rangle_\lambda$ indicates the average measured in a simulation with the Hamiltonian of Eq. \eqref{eq:Hein}, $d_\alpha=\sqrt{2\pi\sigma^2/\beta\alpha}$ is the typical displacement of a particle in a non-interaction Einstein crystal, and $\beta F_\text{Ein}^\text{rot}/N$ is the orientational contribution to the free energy of a non-interacting Einstein crystal. The latter can be calculated by integrating its partition function over all orientations of the particle, i.e.
\begin{align} 
    \frac{\beta F_\text{Ein}^\text{rot}}{N} =&  - \log  \left\{  \frac{1}{8\pi^2}  \int  \exp\left[ -\frac{\beta\alpha}{2} (2 - (\mathbf{\hat{u}}_z\cdot\mathbf{\hat{z}})^2 \right.\right. \nonumber \\
    & \left.\left. - \max_k \{(\mathbf{\hat{u}}_k\cdot\mathbf{\hat{y}})^2 \} ) \right] \sin\theta d\phi d\theta d\psi \right\}  +\log2 , \label{eq:Feinrot}
\end{align}
where $\mathbf{\hat{u}}_z$ and $\mathbf{\hat{u}}_k$ depend on the three Euler angles $\theta$, $\phi$, and $\psi$. Note that the final term, i.e. the $\log2$, is added to Eq. \eqref{eq:Feinrot} to compensate for the difference in rotational freedom caused by using the square of the product $\mathbf{\hat{u}}_k^{(i)}\cdot\mathbf{\hat{y}}$ instead of the product itself.
We numerically estimate the above integral using a MC simulation of $10^{10}$ randomly generated orientations. 
The numerical integration in Eq. \eqref{eq:Feintotal} is performed using a 10-point Gauss-Legendre quadrature \cite{frenkelbook}, and we estimate the error using an additional 11 points from the Gauss-Kronrod rule.


\subsection{Nucleation barrier}
To compute the nucleation barriers, we use MC simulations of $N=10752$ particles in the $NPT$ ensemble combined with umbrella sampling \cite{torrie1974monte,tenwolde1996simulation}. This method adds a biasing potential to the system which drives the simulation towards a preferred region in phase space that is otherwise improbable to be sampled. 
We use the biasing potential \cite{filion2010crystal,dejager2022crystal}
\begin{equation} 
     U_\text{bias}\left(n(\mathbf{r}^N) \right) = \frac{\kappa}{2} \left( n(\mathbf{r}^N) -n_c  \right)^2,
\end{equation}
where $\kappa$ is a coupling, $n(\mathbf{r}^N)$ is the size of the largest cluster present in the configuration $\mathbf{r}^N$, and $n_c$ is the target cluster size. 
By performing a series of simulations, each driving the system towards a specific target size for the crystal nucleus, we can obtain information on the nucleation barrier for different `windows'. 
Combining the resulting parts of the barrier obtained from these windows, we can compute the total nucleation barrier, see e.g. Refs \onlinecite{filion2010crystal,dejager2022crystal}. 
We use isotropic volume changes for these $NPT$ simulations.

In order to measure the nucleus size, we need an order parameter which classifies each particle in the system as either fluid or crystal. 
There are many options for such an order parameter, based on either positional ordering or orientational order, or both.
Note that the specific choice for the order parameter is not important, as the height of the nucleation barrier is independent of any (reasonable) choice of order parameter \cite{filion2010crystal}.
Here, we use an order parameter based on positional ordering; however, in the Supplementary Material we show that using an order parameter based on orientational ordering results in similar classifications for this system.

We capture positional ordering via Steinhardt's bond-orientational order parameters  \cite{steinhardt1983bond}
\begin{equation} \label{eq:qlm}
    q_{lm}(i) = \frac{1}{N_b(i)} \sum_{j\in\mathcal{N}_b(i)} Y_{lm}( \mathbf{r}_{ij}),
\end{equation}
where $N_b(i)$ is the number of neighbors of particle $i$, $\,\mathcal{N}_b(i)$ is the set of neighbors of $i$, $\,Y_{lm}(\mathbf{r}_{ij})$ are the spherical harmonics with $m\in[-l,l]$, and $\mathbf{r}_{ij}=\mathbf{r}_j-\mathbf{r}_i$ is the distance vector between the center of masses of particles $i$ and $j$. 
Using $q_{3m}$, we compute, for each particle pair, the 3-fold symmetric Ten Wolde bond \cite{tenwolde1996numerical}
\begin{equation} \label{eq:d3}
    d_3(i,j) = \frac{ \sum_m q_{3m}(i) q_{3m}^\dagger(j) }{ \sqrt{ \left(\sum_m |q_{3m}(i)|^2 \right) \left(\sum_m |q_{3m}(j)|^2 \right) } },
\end{equation}
where $^\dagger$ indicates the complex conjugate, and $\sum_m$ indicates the sum over $m\in[-3,3]$.
Particle $i$ is then classified as crystal if it has 5 or more neighboring particles $j$ with which it has a crystal-like bond, i.e. $|d_3(i,j)|>0.7$. The neighbors of particle $i$ are defined as all particles $j$ with $|\mathbf{r}_{ij}|<1.1\sigma$, which, for all systems studied, roughly corresponds to the first minimum of the radial distribution function.
Notice that we take the absolute value of $d_3$ to define a bond as crystal-like. This is needed as neighboring particles in the honeycomb lattice have a value for $d_3$ either close to $1.0$ or $-1.0$ (see Supplementary Material). 


\subsection{Onset of nucleation}
To investigate the onset of homogeneous nucleation, we obtain numerous spontaneous nucleation events using brute force MC simulations of $N=10752$ particles in the $NVT$ ensemble.
During these simulations, the configuration of the system is frequently saved, such that we have a detailed time evolution of each nucleation event which can be analyzed in the post processing.
For all events, we search for a precursor, by studying the time evolution of continuous measures for the local density, positional ordering, and orientational ordering in the system. 
The local packing fraction of each particle is computed using the volume of its Voronoi cell, which are obtained using \texttt{voro++} \cite{rycroft2009voro}. 
The local positional ordering is measured using the rotationally invariant bond-orientational order parameters (BOPs) \cite{steinhardt1983bond}
\begin{equation} \label{eq:ql}
    q_l(i) = \sqrt{\frac{4\pi}{2l+1} \sum_m |q_{lm}(i)|^2},
\end{equation}
with $q_{lm}$ as defined in Eq. \eqref{eq:qlm}. 
We use the BOPs with $l\in[1,12]$. 
Note that we chose to use Steinhardt's BOPs instead of an averaged version like the one of Lechner and Dellago \cite{lechner2008accurate}. Steinhardt's BOPs not only provide a more local measure for the positional order, we also found that they are better than Lechner and Dellago's BOPs at distinguishing fluid-like environments from crystalline environments for this system.

To obtain the local orientational order, we use the nematic order parameter
\begin{equation} \label{eq:s2}
    S_2(i) = \frac{1}{N_b(i)} \sum_{j\in\mathcal{N}_b(i)} \left(\frac{3}{2} (\mathbf{\hat{u}}_z^{(i)}\cdot\mathbf{\hat{u}}_z^{(j)})^2 - \frac{1}{2}\right),
\end{equation}
which quantifies the alignment of the long axis of triangular prism $i$ with those of its neighbors $j$.
Although we use a radial cutoff to determine the nearest neighbors for the crystal--fluid classification, we use the SANN algorithm \cite{van2012parameter} to determine the nearest neighbors for the calculation of the BOPs and $S_2$.

Similar to Refs. \onlinecite{berryman2016early,dejager2023search}, 
we track the local density, positional ordering, and orientational ordering at the birthplace of a crystal nucleus over time and look for a head start or delay in the increase of any of the order parameters in comparison to the others. 
For each nucleation event, we first identify the position $\mathbf{r}_0$ that best captures the center of the nucleus at the start of nucleation. We use the average center-of-mass of the precritical nucleus as a starting point for $\mathbf{r}_0$ and, if needed, by eye adjust it to best capture the birthplace of the crystal nucleus. 
Next, starting well before the onset of nucleation, we determine the average local properties of all particles inside a sphere or radius $R$ around $\mathbf{r}_0$ for each configuration.
As we are searching for local precursors, we choose $R$ such that the selected region contains around 30-40 particles. This size provides a good balance between being large enough to obtain relatively stable averages of the local properties, and small enough to ensure that the averaged properties still represent the local situation. 
Somewhat smaller or larger values for $R$, e.g. such that the region contains around 20 or 50 particles, do not significantly influence the trends we observe.

\begin{figure}[t!]
\begin{tabular}{l}
     a) \\[-0.4cm]
     \includegraphics[width=\figwidthA]{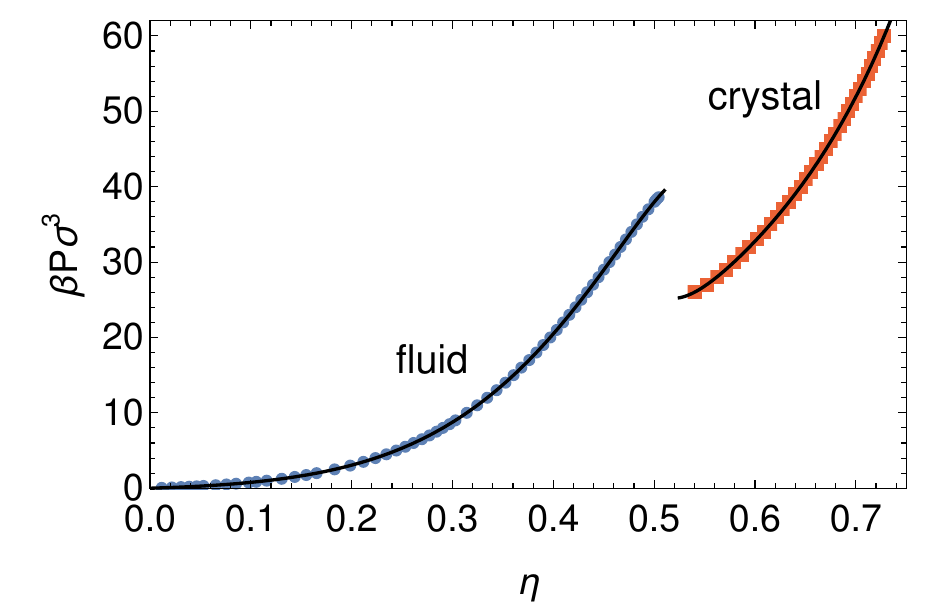} \\
     b) \\[-0.4cm]
    \includegraphics[width=\figwidthA]{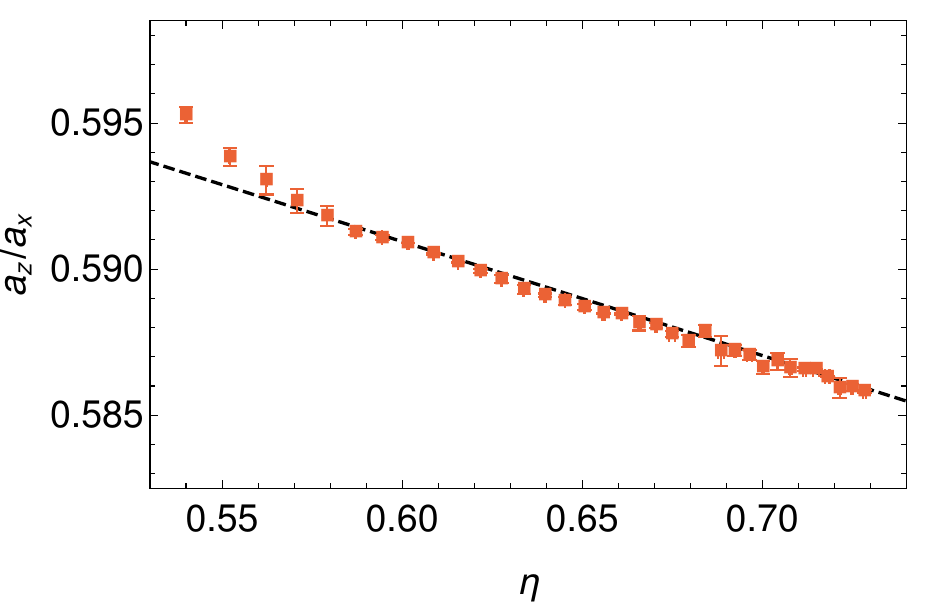} 
\end{tabular}
    \caption{\label{fig:eqstate} 
    \textbf{a)} The equation of state for hard triangular prisms. The errors in the data points are smaller than the markers. The black lines indicate the fits for both the fluid and the crystal phase.
    \textbf{b)} The equilibrium height ratio $a_z/a_x$ of the honeycomb crystal as a function of the packing fraction. The dashed line is a linear guide to the eye.
    }
\end{figure}


\section{Results}


\subsection{Phase behavior}

We start by obtaining the equation of state of triangular prisms using $NPT$ MC simulations. 
The resulting equation of state is shown in Fig. \ref{fig:eqstate}a. 
Like Agarwal and Escobedo \cite{agarwal2011mesophase}, we observe a clear first-order phase transition between the isotropic fluid and the honeycomb crystal. 
In addition, in Fig. \ref{fig:eqstate}b, we show the equilibrium height ratio $a_z/a_x$ for the honeycomb crystal as a function of the packing fraction. Although the effect of the packing fraction is minimal, we observe an approximately linear dependence of the height ratio $a_z/a_x$ on the packing fraction for $\eta\gtrsim0.59$. For smaller packing fractions, the height ratio deviates from this linear trend. This is caused by the fact that these packing fractions lie below the melting point of the crystal, as we will see.

\begin{figure*}[t!]
    \begin{tabular}{cc}
    \includegraphics[width=0.39\linewidth,valign=m]{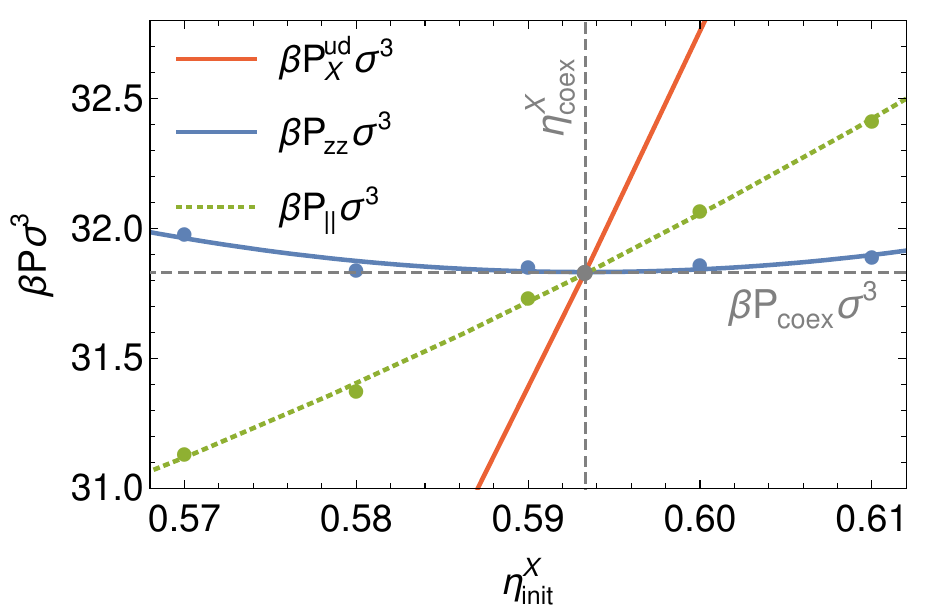} &
    \includegraphics[width=0.59\linewidth,valign=m, raise=0.35cm]{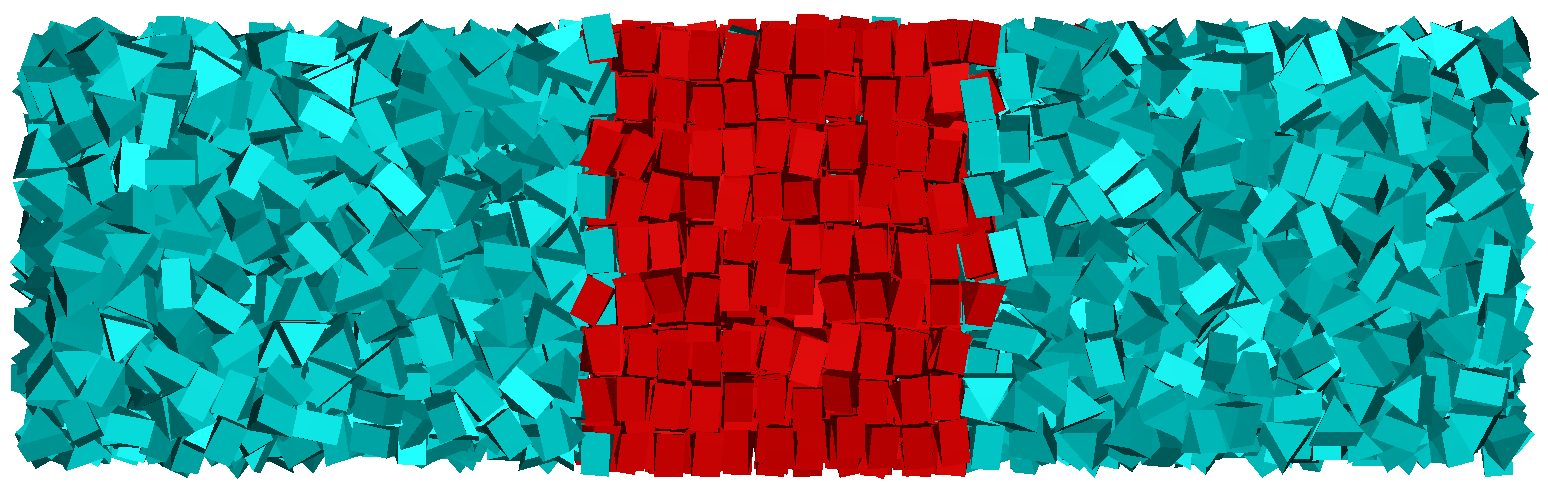} 
    \end{tabular}
    \caption{ \label{fig:directco}
    Direct coexistence approach for a system of $N=4000$ triangular prisms. 
    The plot shows the global pressure $P_{zz}$ normal to the interface as a function of the lattice spacing of the initial crystal $\eta^X_\text{init}$ (blue line). The coexistence point (gray dot) is determined as the crossing point of this line with the bulk equilibrium equation of state (red line). 
    The green dotted line gives the pressure parallel to the interface.
    Statistical errors are on the order of the typical deviations of the points from the fitted lines. 
    The snapshot shows a typical configuration from the direct coexistence simulation. 
    Particles identified as crystal are colored red, and particles identified as fluid are colored blue.
    }
\end{figure*}

In order to determine the boundaries of the phase transition, we fit the equations of states using a 10th order polynomial for the fluid branch and a 6th order polynomial for the crystal branch (black lines in Fig. \ref{fig:eqstate}a). 
Next, we integrate them to determine the free energies of the fluid and crystal phases. For the crystal, we use Einstein integration with finite-size corrections to obtain a reference free energy of a honeycomb crystal at packing fraction $\eta=0.60$ with $a_z/a_x=0.590958$. We use $\alpha=3000k_BT$ as spring constant for the Einstein crystal.
A common tangent construction to the free energies results in the freezing and melting packing fractions $\eta_F=0.4658(3)$ and $\eta_M=0.5935(4)$ and the coexistence pressure $\beta P_\text{coex}\sigma^3=31.86(4)$. The errors are estimated by varying the chosen integration path, e.g. changing the reference packing fraction of the crystal, varying the polynomial order of the fits, and using different spring constants. 

Note that the coexistence pressure is slightly lower than the one reported by Agarwal and Escobedo\cite{agarwal2011mesophase}, i.e. $\beta P_\text{coex}\sigma^3=32.7(1)$ (with $\eta_F=0.47$ and $\eta_M=0.59$). 
Hence, to further validate our coexistence pressure, we additionally perform direct coexistence simulations. Following Ref. \onlinecite{smallenburg2024simple}, we perform event-driven molecular dynamics (EDMD) simulations of $N=4000$ triangular prisms in the $NVT$ ensemble. 
For these EDMD simulations, we use the code of Ref. \onlinecite{smallenburg2022efficient}, adapted to simulate anisotropic particles\cite{van2017phase} and measure the global pressure tensor. 
The periodic simulation box is elongated in the $z$-direction and contains a direct coexistence between the fluid and crystal phases, with the two interfaces perpendicular to the long axis of the box. See Fig. \ref{fig:directco} for an example snapshot of the system.
To find the coexistence conditions, we search for the initial crystal packing fraction, $\eta^X_\text{init}$, for which the crystal phase is unstrained. 
In practice, this happens when the global pressure normal to the interface is equal to the pressure of an undeformed bulk crystal at the same packing fraction $\eta^X_\text{init}$, i.e. $P_{zz}=P_X^\text{ud}$. 
We consider a small range of initial packing fractions $\eta^X_\text{init}\in[0.57,0.61]$, and show the resulting global pressures $P_{zz}$ in Fig. \ref{fig:directco}. 
We fit $P_{zz}$ using a quadratic equation, and from its crossing point with $P_X^\text{ud}$ we obtain the coexistence pressure $\beta P_\text{coex}\sigma^3=31.83(3)$, which is in excellent agreement with our value obtained from free-energy calculations.
Considering the behavior of the pressures in Fig. \ref{fig:directco}, it is not unreasonable to assume that the slightly higher coexistence pressure of Agarwal and Escobedo may be due to some strain in their direct coexistence simulations.
Lastly, notice that, at coexistence, the pressure parallel to the interface, $P_\parallel=(P_{xx}+P_{yy})/2$, is essentially equal to $P_{zz}$. While this condition is not a requirement for coexistence, it indicates that the surface stress $f$ of this interfacial plane is zero, as \cite{dejager2024thermodynamics}
\begin{equation} \label{eq:stress}
    f_ = \frac{1}{2} L_z \left( P_{zz} - P_\parallel \right), 
\end{equation}
where $L_z$ is the length of the box in the $z$-direction.


\subsection{Nucleation barrier}

\begin{table}[t!]
\caption{\label{tab:info} 
Information on the two systems studied for their crystal nucleation.
The last three columns give the interfacial free energy, the barrier height, and critical nucleus size, all obtained from fitting the nucleation barrier to CNT. 
The error in $\beta\Delta G^*$ is no more than 1. 
}
\begin{ruledtabular}
\begin{tabular}{ccccccc}
    $\eta$ & $\beta P\sigma^3$ & $\eta_X$ & $\beta |\Delta\mu|$ & $\beta\gamma\sigma^2$ & $\beta\Delta G^*$ & $n^*$ \\ \hline
    0.501  & 38.1  & 0.634  & 0.69  & 1.40  & 15.9  & 30  \\ 
    0.505  & 38.7  & 0.638  & 0.76  & 1.21  & 13.0  & 23  \\
\end{tabular}
\end{ruledtabular}
\end{table}

\begin{figure}[t!]
     \centering
     \includegraphics[width=\figwidthA]{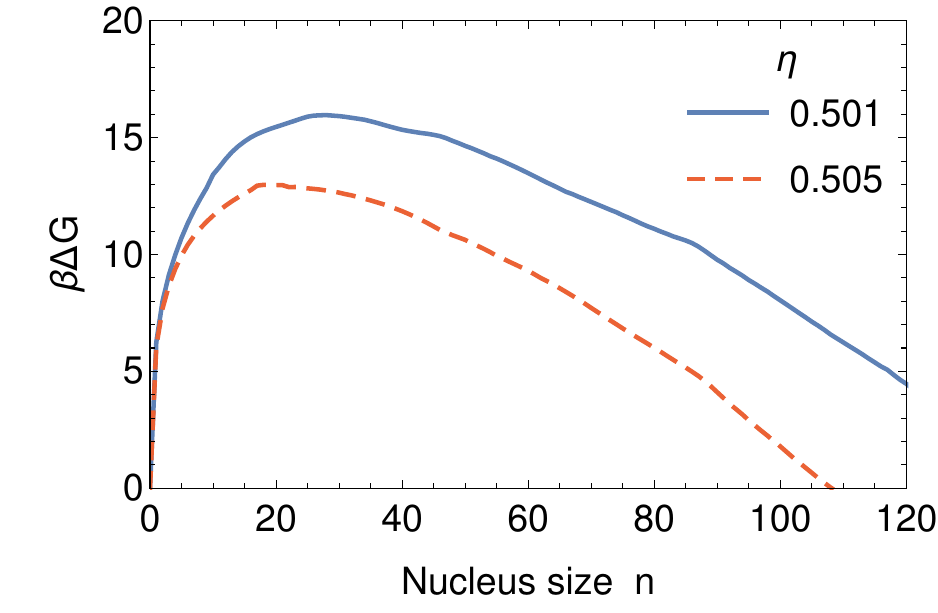}
     \caption{\label{fig:barrier}
     The Gibbs free-energy barrier as a function of the nucleus size for both supersaturations studied (see Tab. \ref{tab:info}).
     }
\end{figure}

We now turn our attention to the crystal nucleation of hard triangular prisms. In this work, we focus on two supersaturations: a metastable fluid at packing fraction $\eta=0.501$ and at $\eta=0.505$. 
The corresponding pressures, packing fractions of the crystal, and supersaturations are given in Tab. \ref{tab:info}. 
Using umbrella sampling with a biasing strength of $\kappa=0.1k_BT$, we perform $NPT$ MC simulations for target nucleus sizes $n_c\in[20,120]$ with an interval of 10. For each window, we perform 3 independent simulations. The resulting nucleation barriers are shown in Fig. \ref{fig:barrier}. By fitting approximately the top third of the nucleation barrier to CNT (see e.g. Refs. \onlinecite{filion2010crystal,dejager2022crystal}), we obtain the barrier height, critical nucleus size, and interfacial free energy, which are also reported in Tab. \ref{tab:info}.
We note that the heights of these nucleation barriers are reasonable for the system to overcome during brute force simulations, while simultaneously being high enough such that each simulation will likely only contain a single (successful) nucleation event. 
Furthermore, contrary to hard cubes \cite{sharma2021low}, we find that the hard triangular prisms exhibit reasonable high interfacial free energies. Consequently, we expect a more classical nucleation pathway, as observed for hard spheres \cite{berryman2016early,dejager2023search}, instead of the non-classical pathway observed for hard cubes \cite{sharma2018disorder}.

\subsection{Onset of nucleation}

 \begin{figure}[t!]
     \centering
     \includegraphics[width=\figwidthA]{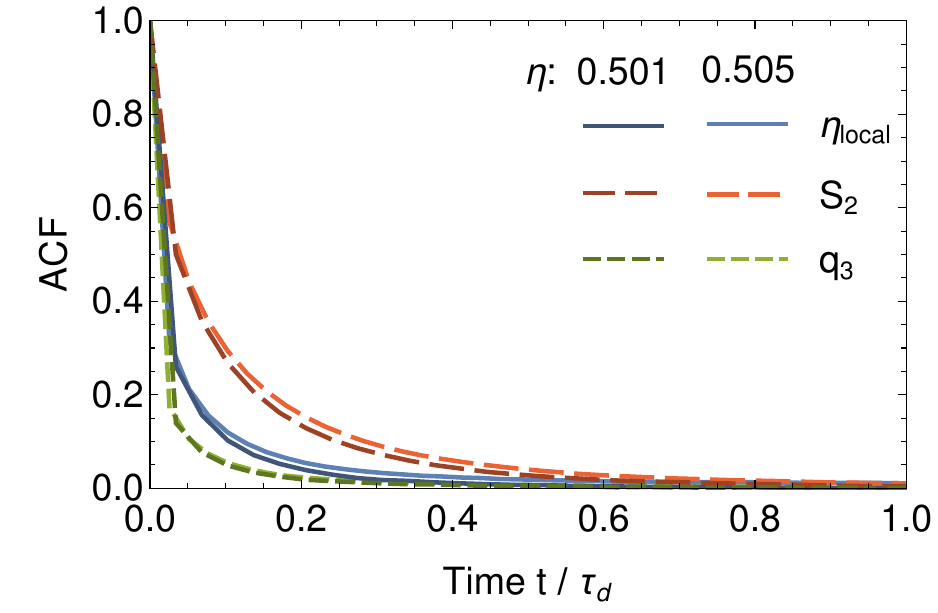}
     \caption{\label{fig:acf}
     The autocorrelation function of the local packing fraction, nematic order parameter, and  3-fold symmetric BOP in the metastable fluids of packing fraction $\eta=0.501$ (light) and $\eta=0.505$ (dark).}
 \end{figure}

To investigate the onset of homogeneous nucleation, we focus again on metastable fluids at packing fractions $\eta=0.501$ and $\eta=0.505$. 
We perform brute force $NVT$ MC simulations to generate spontaneous nucleation events, and study the time evolution of local properties of the region in which the nucleus is born.
However, in order to better interpret the temporal fluctuations of these local properties, we first determine their autocorrelation functions (ACFs) inside these metastable fluids. 
Figure \ref{fig:acf} shows the ACF of the local packing fraction, nematic order parameter, and the 3-fold symmetric BOP. 
Here, we give the time in terms of the long-time translational diffusion time $\tau_d=\sigma_c^2/6D_l$, where $D_l$ is the long-time translational diffusion coefficient obtained from the mean-squared displacement (in the fluid) and $\sigma_c=\sqrt{5/3}\sigma$ is the diameter of the circumscribed sphere of the triangular prism.
Notice that the specific ACFs for both fluids are very similar, and that all decay extremely fast, i.e. within half a diffusion time. We find that the nematic order parameter has the longest lived correlations, followed by the ACF of the local packing fraction. Although $q_3$ clearly has the shortest lived correlations out of these three parameters, it has the longest decay time of all BOPs. 
The decay time of these parameters provides us with a rough time window prior to the start of nucleation in which we can search for a precursor. 
As the nematic order parameter has the longest decay time, we will focus on the time window for which $\text{ACF}(S_2)>0.05$.

\newcommand{\figwidthS}{0.24\linewidth}
\begin{figure*}[t!]
\begin{tabular}{llll}
	a) $\;\quad(t-t_0)/\tau_d=-0.07$ & b) $\;\quad(t-t_0)/\tau_d=1.99$ & c) $\;\quad(t-t_0)/\tau_d=5.62$ & d) $\;\quad(t-t_0)/\tau_d=10.91$ \\
	\includegraphics[width=\figwidthS]{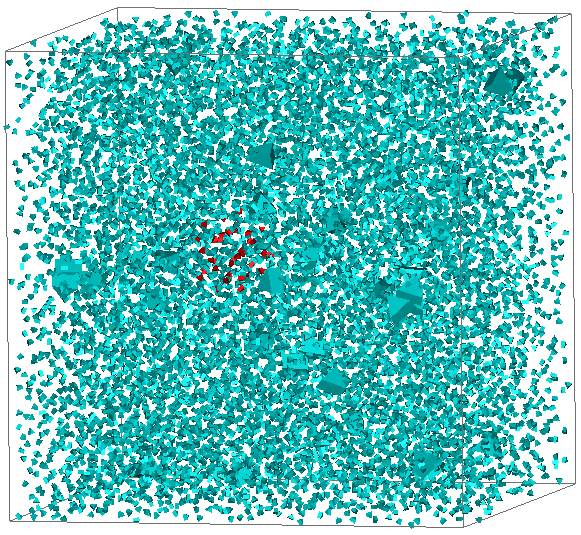}
	& \includegraphics[width=\figwidthS]{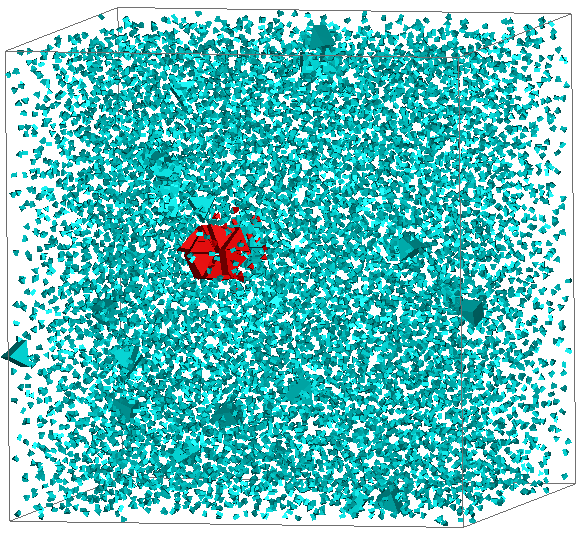}
	& \includegraphics[width=\figwidthS]{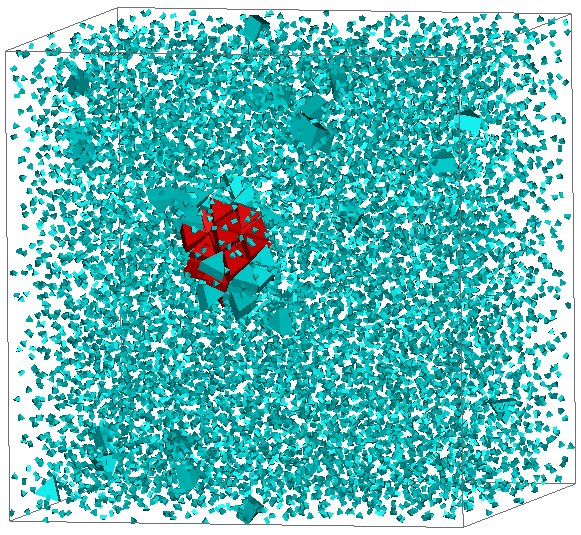}
	& \includegraphics[width=\figwidthS]{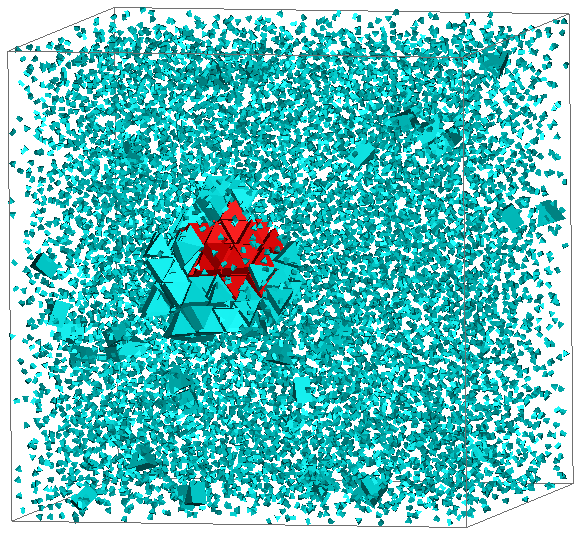}
	\end{tabular}
	\caption[width=1\linewidth]{\label{fig:selectregion} 
    \textbf{(a-d)} Short time series of a typical nucleation event ($\eta=0.501$). Here $t_0$ indicates the start of nucleation and time is given in terms of the long-time diffusion time $\tau_d$. Fluid particles are displayed at a quarter of their actual size to make the nucleus visible, and red indicates the particles inside the studied region, i.e. those inside the sphere of radius $R$ around the center of nucleation $\mathbf{r}_0$. 
	}
\end{figure*}

\begin{figure*}[t!]
    \centering
    \begin{tabular}{cc}
        \includegraphics[width=\figwidthB]{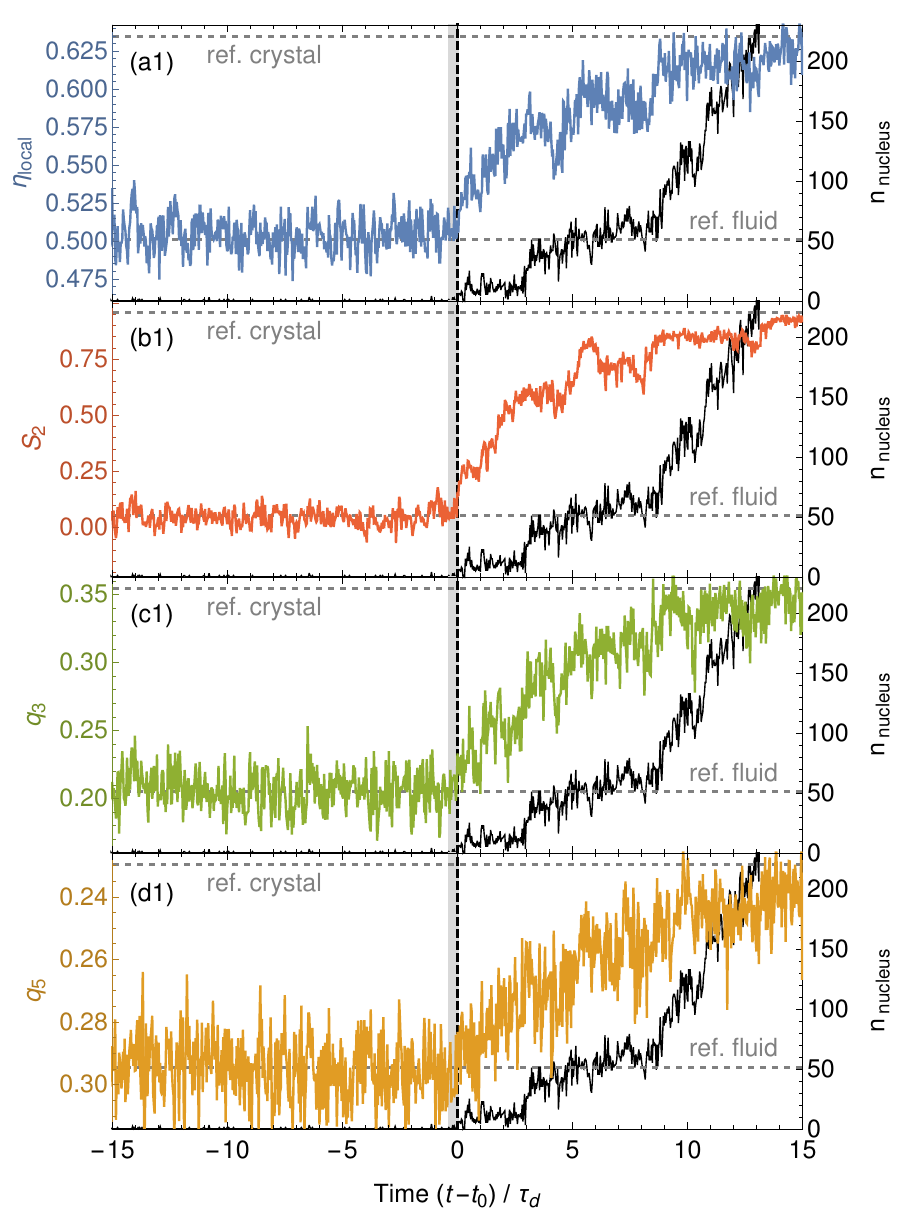}
        & \includegraphics[width=\figwidthZ]{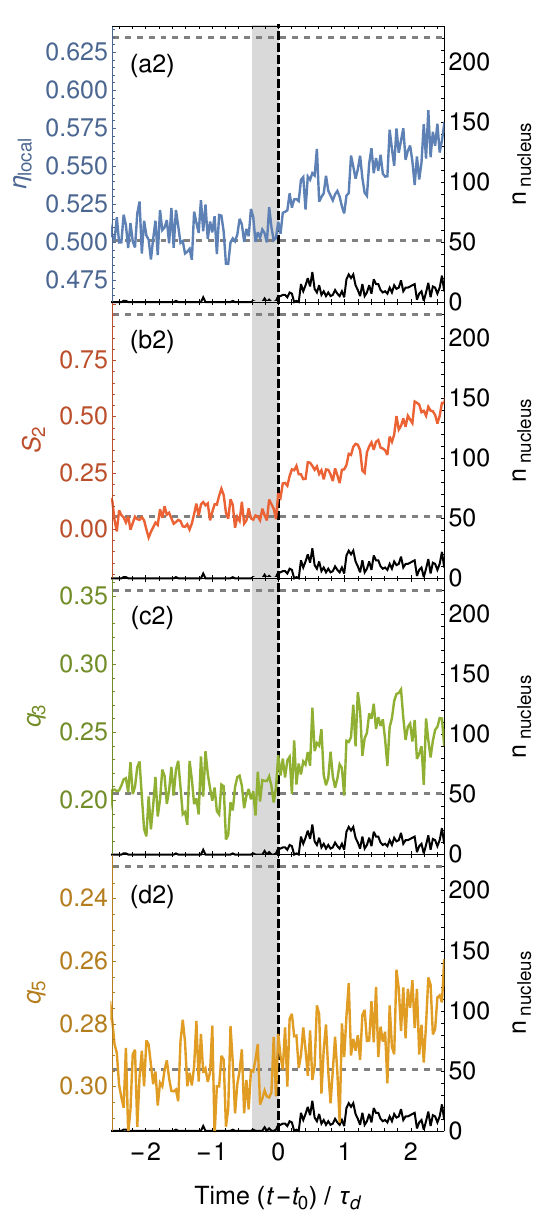}
    \end{tabular}
    \caption[width=1\linewidth]{\label{fig:event501} 
    Time evolution of the average local order parameters at the birthplace of the crystal nucleus for a typical nucleation event at packing fraction $\eta=0.501$. 
    This is the same event as in Fig. \ref{fig:selectregion}.
    The vertical dashed line in each figure indicates the start of nucleation $t_0$, and the shaded area indicates the time window before $t_0$ for which $\text{ACF}(S_2)>0.05$ (see Fig. \ref{fig:acf}). The left panels, \textbf{(a1-d1)}, show the evolution of the parameters over a long time interval, and the right panels, \textbf{(a2-d2)}, show the same evolutions but now zoomed in on the short time interval around the start of nucleation. In each figure, the black line indicates the number of particles in the crystal nucleus. The colored lines indicate \textbf{(a)} the local packing fraction, \textbf{(b)} the nematic order parameter, and the \textbf{(c)} 3-fold and \textbf{(d)} 5-fold symmetric BOPs. 
    Note that the vertical axis of \textbf{(d)} is inverted such that, even though $q_5$ is decreasing as nucleation progresses, the inverted trend shows a similar increase as for the other order parameters.
    The horizontal dashed lines indicate the reference value of each parameter in the bulk fluid and crystal phases. 
    The left vertical axis of each figure is scaled in such a way that they all depict the same relative range with respect to these bulk values.
    }
\end{figure*}

\begin{figure*}[t!]
    \centering
    \begin{tabular}{cc}
        \includegraphics[width=\figwidthB]{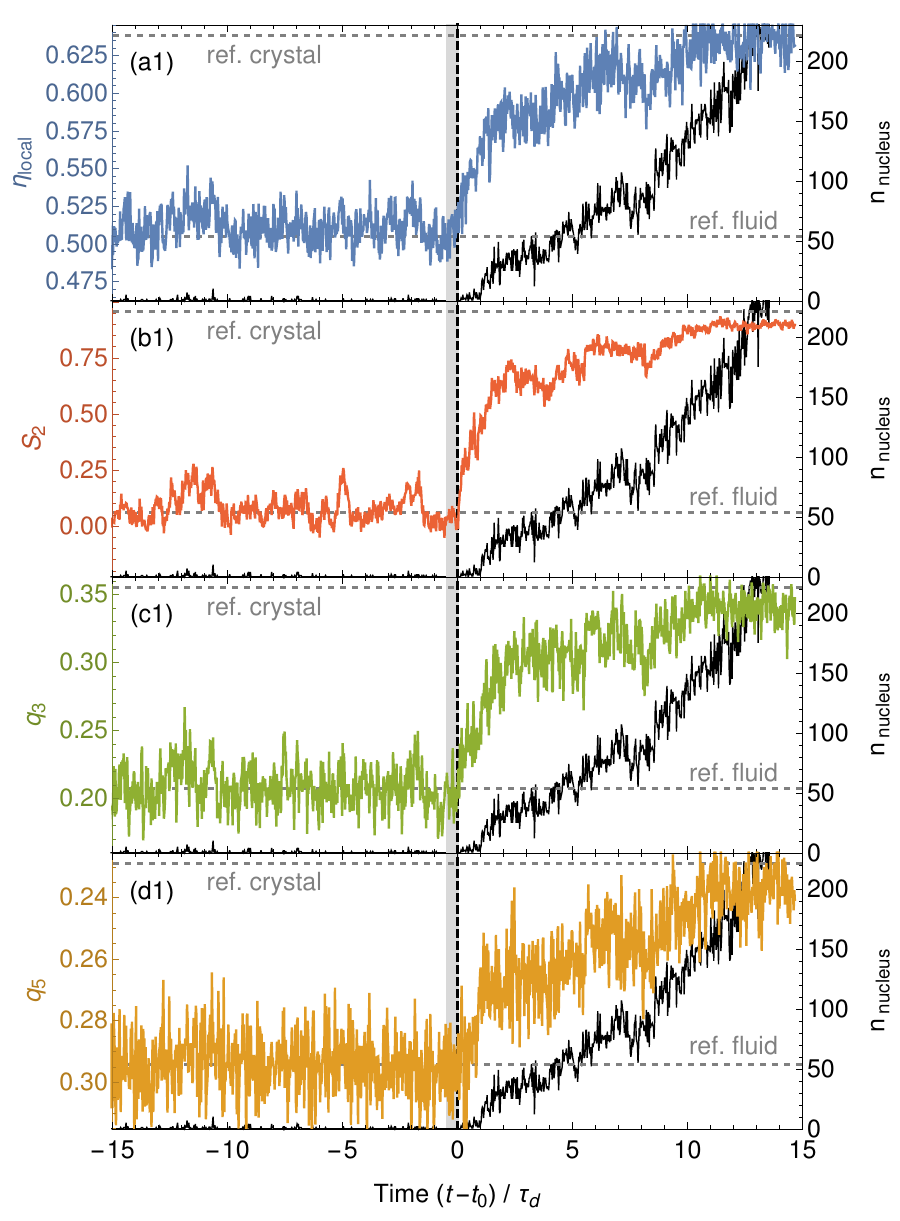}
        & \includegraphics[width=\figwidthZ]{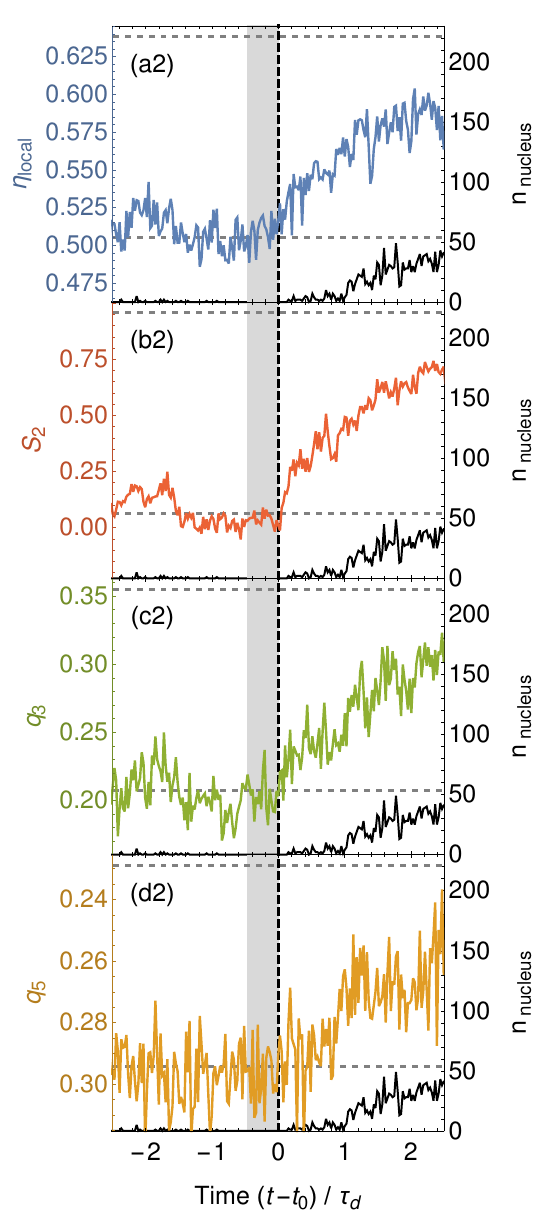}
    \end{tabular}
    \caption[width=1\linewidth]{\label{fig:event505} 
    Time evolution of the average local order parameters at the birthplace of the crystal nucleus for a typical nucleation event at packing fraction $\eta=0.505$. 
    See caption of Fig. \ref{fig:event501} for an explanation of the figures.}
\end{figure*}

In total, we study 18 nucleation events for $\eta=0.501$ and 30 events for $\eta=0.505$.
Even though the triangular prisms are highly anisotropic, all events exhibit the spontaneous formation (and growth) of a roughly spherical crystal nucleus. 
This is not a priori a given for anisotropic particles, e.g. gyrobifastigia form egg-shaped crystal nuclei \cite{sharma2018nucleus}.
For each nucleation event, we determine the center of nucleation $\mathbf{r}_0$, and monitor the properties of the particles with $|\mathbf{r}(t)-\mathbf{r}_0|<R$, where $R=1.65\sigma$, over time. 
To better illustrate which particles are included in this spherical region, Fig. \ref{fig:selectregion} shows a short time series of a typical nucleation event in which the particles inside the studied region are colored in red. 
Note that $\mathbf{r}_0$ is chosen to best capture the start of nucleation and is kept fixed for the entire time series. As a consequence, the center of mass of the nucleus does not necessarily coincide with $\mathbf{r}_0$ at every time step.
As it is impractical to average the time evolution of the order parameters of multiple nucleation events together, we will discuss our results and observations using two typical events as examples. 
However, the observations that we make are based on the results of all studied events.
In order to demonstrate the legitimacy of our observations, the Supplementary Material contains the results on some additional nucleation events.

The time evolution of the local parameters in the studied region are shown in Figs. \ref{fig:event501} and \ref{fig:event505} for a typical nucleation event at $\eta=0.501$ and $\eta=0.505$, respectively.
Note that the event of Fig. \ref{fig:event501} is the same event of which the snapshots are shown in Fig. \ref{fig:selectregion}.
In addition to the local order parameters, Figs. \ref{fig:event501} and \ref{fig:event505} also show the nucleus size as a function of time. 
Although the nucleus size is unsuitable as an order parameter for monitoring the onset of nucleation, it can provide a general overview of the nucleation event, such as the speed of nucleus growth or the time at which the nucleus reaches its critical size (see Tab. \ref{tab:info}).

The first thing to notice about the nucleation trajectories of Figs. \ref{fig:event501} and \ref{fig:event505} is the extreme narrowness of the time window (indicated in gray) prior to the start of nucleation in which we can search for a precursor. Especially in comparison to the time it takes the nucleus to grow to a significant size.
Recall that we obtained this time window from the ACF of the local order parameters,
and that local properties of the fluid are only correlated to those of the fluid at $t=t_0$ inside this time window.
Nevertheless,
for both fluids we observe no notable change in the behavior of any of the order parameters during this time window. 
Note, though, that the metastable fluid at $\eta=0.505$ is in general more noisy and  more prone to large fluctuations in the local order parameters than the metastable fluid at $\eta=0.501$. This is not surprising, as a more supersaturated fluid simply has a higher affinity for the spontaneous crystallization and melting of tiny nuclei than the less supersaturated fluid.

Furthermore, we observe that, once nucleation starts, it does so abruptly with a strong and simultaneous change in all order parameters. Not only do we observe a simultaneous increase in the local density and local positional ordering, as was likewise observed for hard and charged spheres \cite{dejager2023search}, the orientational ordering of these anisotropic particles increases hand in hand with local density as well. 
Notice that we inverted the vertical axis of the positional order parameter $q_5$, such that its inverted trend increases, 
even though it actually decreases during nucleation and growth.
Moreover, we observe that all order parameters of an event follow roughly identical trends while increasing (or decreasing in the case of $q_5$). 
This is made apparent by our deliberate scaling of the left vertical axis of each figure, such that they all show the same range relative to the reference values of that specific order parameter in the bulk crystal and fluid phases.
In the Supplemental Material, we show for some additional BOPs that they roughly follow these same trends as well.

All these observations combined strongly indicate that local densification, positional ordering, and orientational ordering go together, and, thus, that there is no apparent precursor.


\section{Conclusion}
To conclude, we have determined the phase boundaries of hard triangular prisms, and computed the crystal nucleation barriers for two metastable fluids. 
Furthermore, we have investigated the onset of crystal nucleation in the same two metastable fluids.
By monitoring the time evolution of the local density and positional and orientational ordering in the spatial region coinciding with the birthplace of a crystal nucleus, we demonstrated that all local order parameters increase simultaneously as soon as nucleation starts. Furthermore, we observed no (consistent) atypical behavior in any of the local order parameters prior to the start of nucleation. 
All our results demonstrate that crystal nucleation of hard triangular prisms is a truly spontaneous event in which not only local density and positional ordering increase simultaneously, as for hard and charged spheres \cite{dejager2023search}, but also orientational ordering. 
Thus, we conclude that we find no evidence for a precursor for crystal nucleation of hard triangular prisms.


\section{Supplementary Material}
In the Supplementary Material we provide additional information on the fluid--crystal order parameter, as well as the time evolutions of the local parameters of some additional nucleation events.


\section{Acknowledgements}
We would like to thank Frank Smallenburg and Rinske Alkemade for providing the EDMD code. 
L.F. and M.d.J. acknowledge funding from the Vidi research program with project number VI.VIDI.192.102 which is financed by the Dutch Research Council (NWO).

\section{Data Availability Statement}
An open data package containing the (analyzed) data and other means to reproduce the results of the simulations will be published on Zenodo.

\bibliography{paper}

\end{document}


\preprint{APS/123-QED}

\title{Supplemental Material for ``Phase behavior and crystal nucleation of hard triangular prisms''}

\author{Marjolein de Jager}
\affiliation{Soft Condensed Matter and Biophysics, Debye Institute for Nanomaterials Science, Utrecht University, Utrecht, Netherlands}
\author{Nena Slaats}
\affiliation{Soft Condensed Matter and Biophysics, Debye Institute for Nanomaterials Science, Utrecht University, Utrecht, Netherlands}
\author{Laura Filion}
\affiliation{Soft Condensed Matter and Biophysics, Debye Institute for Nanomaterials Science, Utrecht University, Utrecht, Netherlands}


\maketitle

\onecolumngrid  


\newcommand{\comment}[1]{{\color{red}{\bf #1}}}

\renewcommand{\thetable}{S\arabic{table}}
\renewcommand{\thefigure}{S\arabic{figure}}
\renewcommand{\theequation}{S\arabic{equation}}

\newcommand{\figwidthA}{0.44\linewidth}

\newcommand{\figwidthB}{0.6\linewidth}
\newcommand{\figwidthZ}{0.361\linewidth}


This Supplementary Material includes additional information on the fluid--crystal order parameter discussed in the main paper, as well as the time evolutions of the local parameters of some additional nucleation events.

\section{Order parameter for fluid--crystal classification}
In order to identify the crystal nuclei in our systems, we use an order parameter which classifies a particle as either fluid or crystal. 
In the main paper, we define an order parameter based on the positional ordering of a particle, and use this for the umbrella sampling. 
Here, we first provide some additional information to validate our choices in certain threshold values. Second, we show that an order parameter based on the orientational ordering results in very similar classifications for this system.

\begin{figure*}[h!]
\begin{tabular}{lll}
     (a) & \hspace{0.5cm} & (b)  \\[-0.4cm]
     \includegraphics[width=\figwidthA]{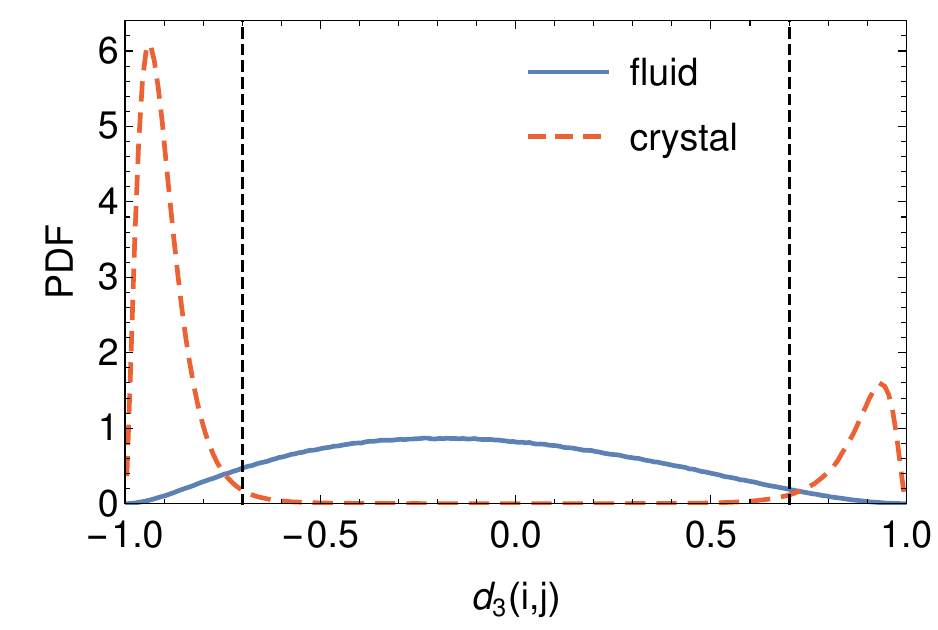} & & \includegraphics[width=\figwidthA]{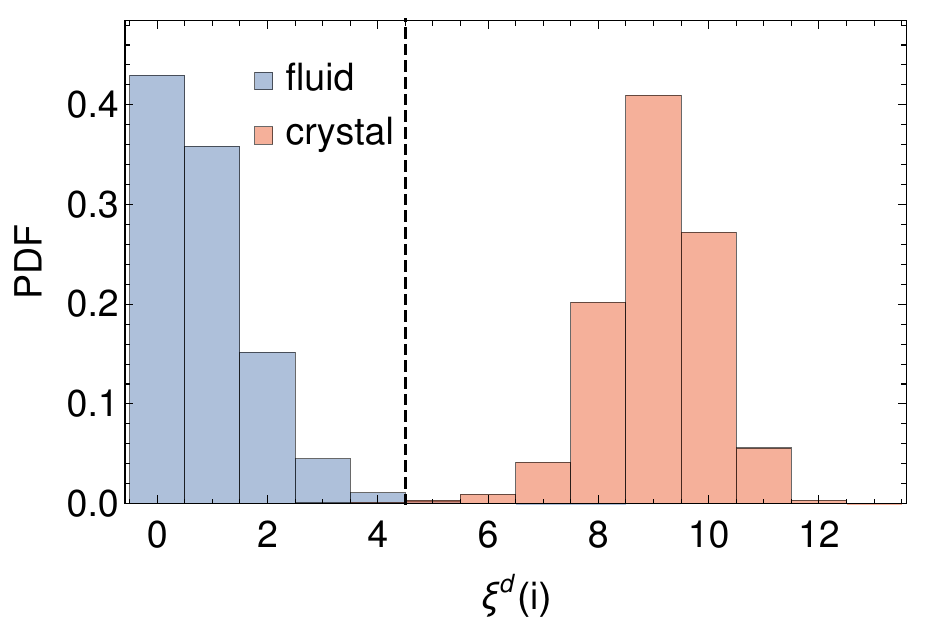} 
\end{tabular}
    \caption[width=1\linewidth]{\label{fig:pdfd3} 
    Probability distribution of \textbf{(a)} the 3-fold symmetric Ten Wolde bond $d_3(i,j)$ and \textbf{(b)} the number of crystal-like bonds per particle (i.e. when $|d_3(i,j)|>d_c$) for the bulk fluid ($\eta=0.501$) and the crystal ($\eta=0.6344$) phases. The vertical dashed lines indicate the cutoff parameters $d_c=0.7$ and $\xi^d_c=5$.}
\end{figure*}

The order parameter used throughout this work is based on the 3-fold symmetric Ten Wolde bond \cite{tenwolde1996numerical} $d_3(i,j)$, which provides a measure for the 3-fold positional symmetry of a bond between particles $i$ and $j$ (see Eq. (9) of main paper). 
In Fig. \ref{fig:pdfd3}a, we show the probability distribution of $d_3(i,j)$ for all pairs of neighboring particles inside a (metastable) fluid phase and a (thermal) honeycomb crystal. We clearly see that $d_3(i,j)$ has a very broad distribution in the fluid, whereas its distribution exhibits well-defined peaks near $\pm1$ in the crystal. We, therefore, chose to classify bonds with $|d_3(i,j)|>0.7$ as crystal-like, and define the parameter $\xi^d$ as the number of crystal-like bonds per particle. 
In Fig. \ref{fig:pdfd3}b, we show the probability distribution of $\xi^d$. We clearly see that the fluid phase mostly contains particles with $\xi^d<5$ and the crystal phase mostly contains particles with $\xi^d>5$. Hence, we chose to classify a particle as crystal when $\xi^d\geq5$.

\begin{figure*}[t!]
\begin{tabular}{lll}
     (a) & \hspace{0.5cm} & (b)  \\[-0.4cm]
     \includegraphics[width=\figwidthA]{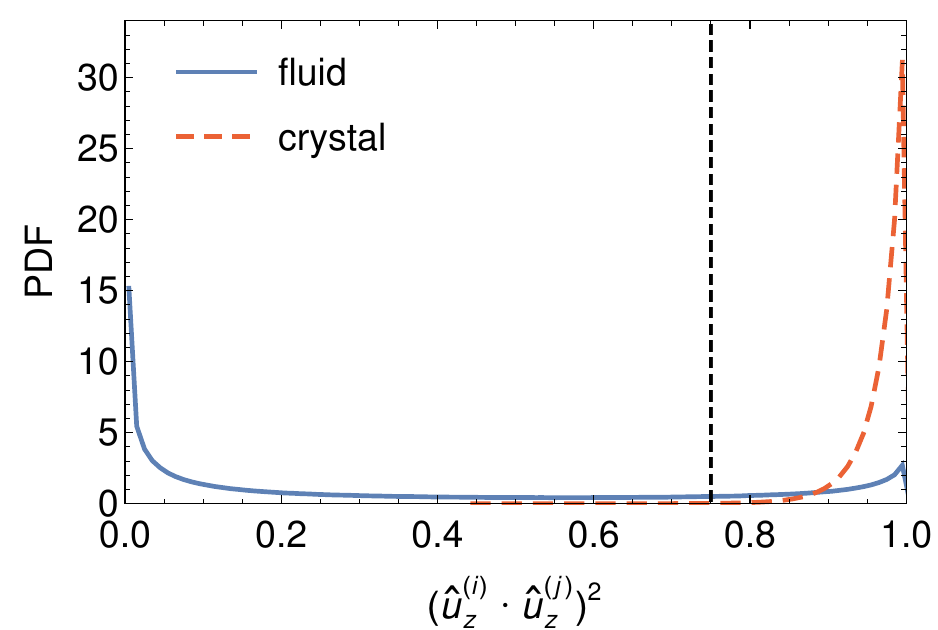} & & \includegraphics[width=\figwidthA]{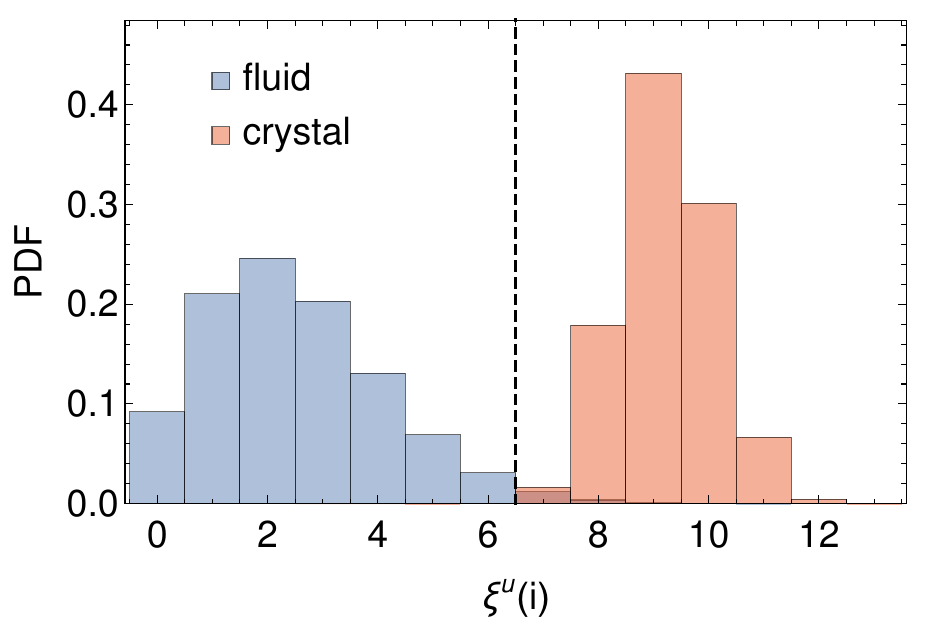} 
\end{tabular}
    \caption[width=1\linewidth]{\label{fig:pdfu} 
    Probability distribution of \textbf{(a)} the long-axis alignment $(\mathbf{\hat{u}}_z^{(i)}\cdot\mathbf{\hat{u}}_z^{(j)})^2$ and \textbf{(b)} the number of crystal-like bonds per particle (i.e. when $(\mathbf{\hat{u}}_z^{(i)}\cdot\mathbf{\hat{u}}_z^{(j)})^2>u_c$) for the bulk fluid ($\eta=0.501$) and the crystal ($\eta=0.6344$) phases. The vertical dashed lines indicate the cutoff parameters $u_c=0.75$ and $\xi^u_c=7$.}
\end{figure*}

Next, we define, in a similar fashion, an order parameter based on the orientational ordering of a particle. We base this order parameter on long-axis alignment of neighboring particles, which is captured by $(\mathbf{\hat{u}}_z^{(i)}\cdot\mathbf{\hat{u}}_z^{(j)})^2$.  
In Fig. \ref{fig:pdfu}a, we show the probability distribution of this value for a (metastable) fluid phase and a (thermal) crystal. The crystal phase has a sharp and well-defined peak at $(\mathbf{\hat{u}}_z^{(i)}\cdot\mathbf{\hat{u}}_z^{(j)})^2\simeq1$. We, therefore, chose to classify bonds between neighboring particles as crystal-like when $(\mathbf{\hat{u}}_z^{(i)}\cdot\mathbf{\hat{u}}_z^{(j)})^2>0.75$, and define the parameter $\xi^u$ as the number of crystal-like bonds per particle. 
In Fig. \ref{fig:pdfu}b, we show the probability distribution of $\xi^u$. We clearly see that the fluid phase mostly contains particles with $\xi^u<7$ and the crystal phase mostly contains particles with $\xi^u>7$. Hence, we chose to classify a particle as crystal when $\xi^u\geq7$.

\vspace{0.4cm}

\clearpage

 \begin{figure}[t!]
     \centering
     \includegraphics[width=0.47\linewidth]{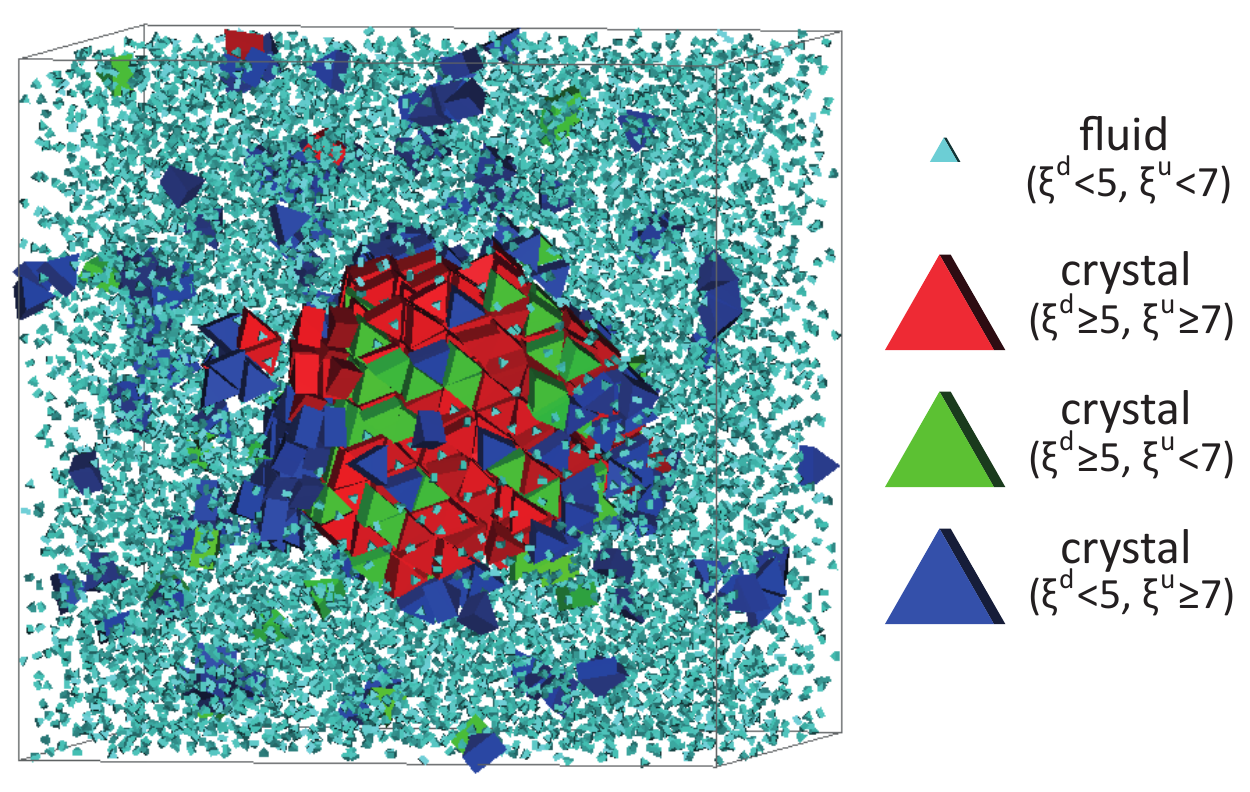}
     \caption{\label{fig:classification}
     Snapshot of a fluid containing a crystal nucleus. Fluid particles are displayed at a quarter of their actual size to make the nucleus visible. Particles classified as crystal by both positional ordering ($\xi^d\geq5$) and orientational ordering ($\xi^u\geq7$) are colored red, particles only classified as crystal by positional ordering are colored green, and particles only classified as crystal by orientational ordering are colored dark blue.
     }
 \end{figure}

In order to compare both classifications, in Fig. \ref{fig:classification} we show a snapshot of a system containing a crystal nucleus and color the particles according to their classification. Particles classified as crystal by both $\xi^d$ and $\xi^u$ are colored red, particles classified as crystal by $\xi^d$ but not by $\xi^u$ are colored green, and particles classified as crystal by $\xi^u$ but not by $\xi^d$ are colored dark blue.
We clearly see that the majority of the nucleus is classified as crystal by both $\xi^d$ and $\xi^u$. Only some surface particles are classified differently by $\xi^d$ and $\xi^u$.

As a last remark, we want to stress that there are many more options for classifying particles as either crystal or fluid. One could also, e.g., use an order parameter which classifies particles based on a certain threshold for $q_3$ or $S_2$. 
However, the choice of order parameter does not influence our results, as any reasonable choice for the order parameter will result in nucleation barriers with the same height \cite{filion2010crystal}.


\section{Additional analysis and nucleation events}
In the main paper, we discuss our observations and results using two typical nucleation events. 
Here, we show for one of these events the time evolution of some additional bond orientational order parameters (BOPs), as well as the results on some additional nucleation events. 
As in the main paper, all figures show the average values of the local order parameters inside the spherical region coinciding with the birthplace of the crystal nucleus. 
In each figure, the vertical dashed line indicates the start of nucleation $t_0$ and the horizontal dashed lines indicate the reference values of that specific order parameter in the bulk crystal and fluid phases. Note that the left vertical axis of each figure is scaled in such a way that they all depict the same relative range with respect to these bulk values.

\begin{figure*}[h!]
    \centering
    \begin{tabular}{cc}
        \includegraphics[width=\figwidthB]{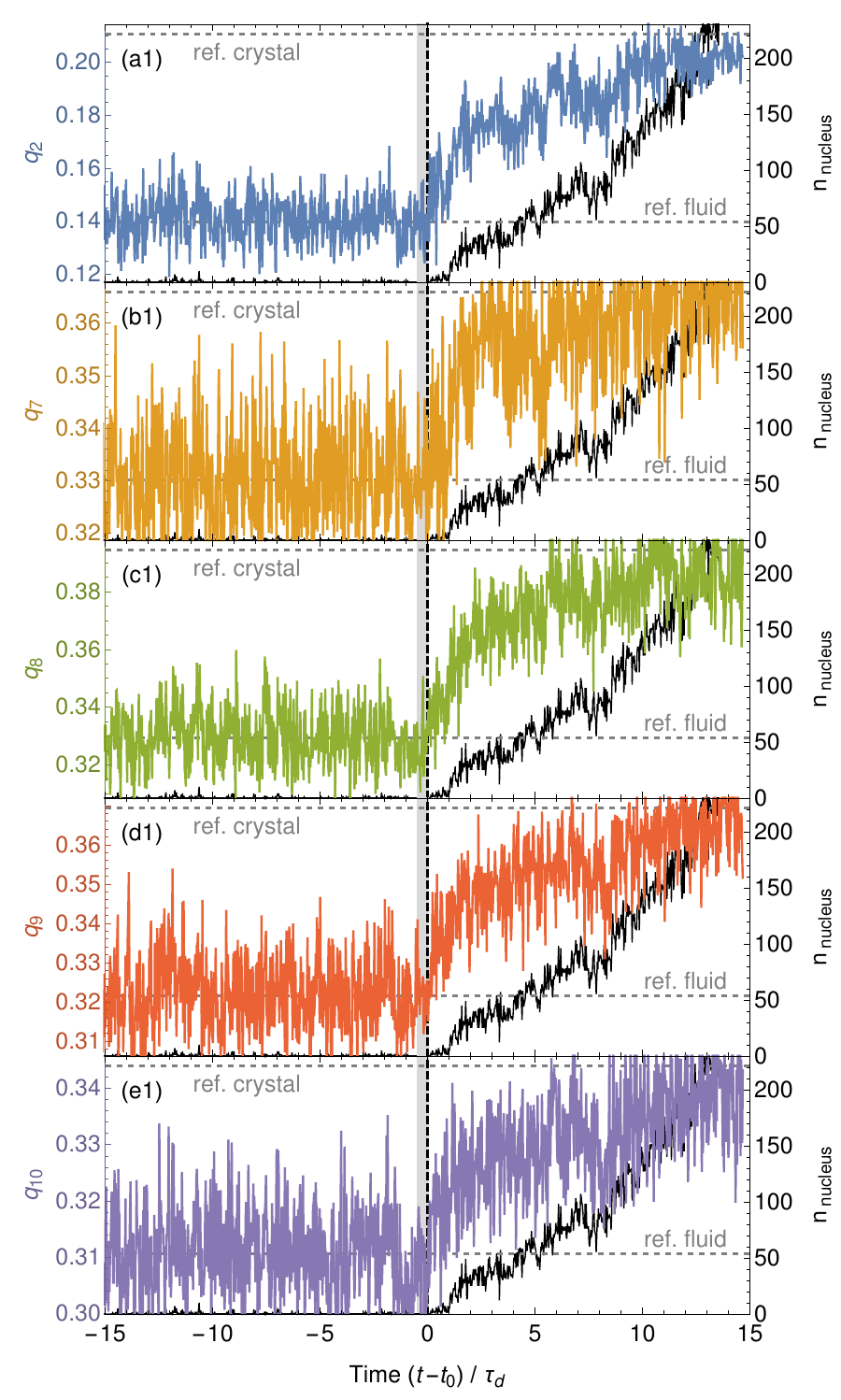}
        & \includegraphics[width=\figwidthZ]{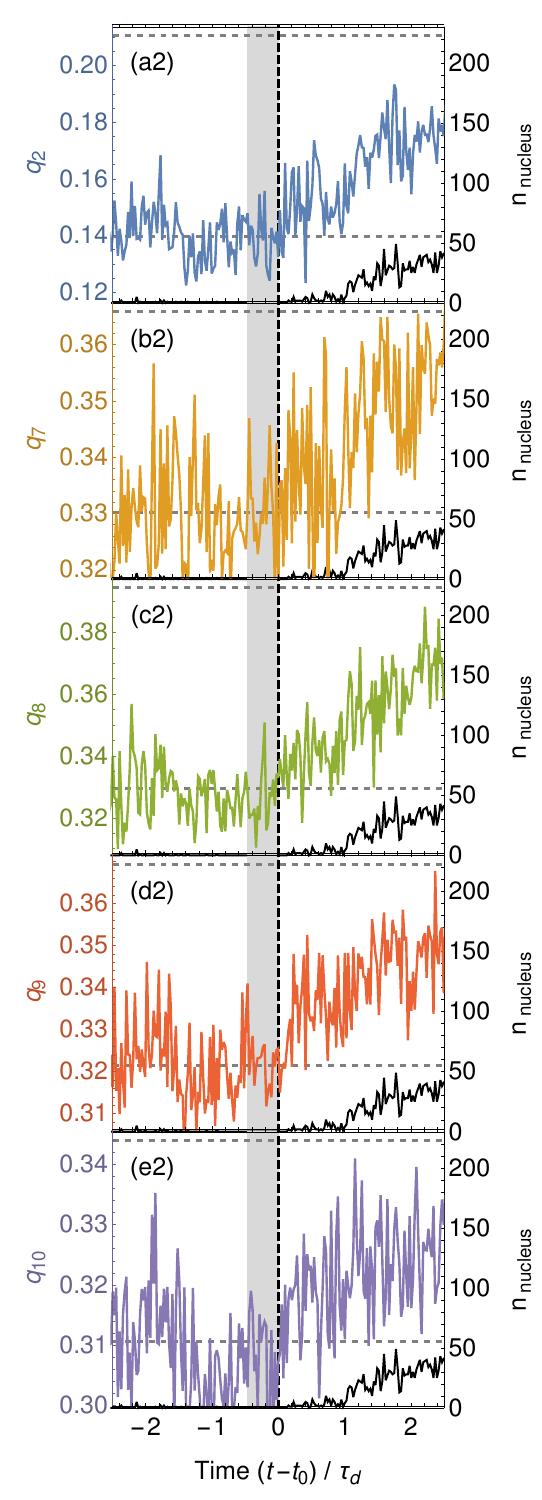}
    \end{tabular}
    \caption[width=1\linewidth]{\label{fig:sievent505moreQ} 
    For the same event as in Fig. 8 of the main paper ($\eta=0.505$), the time evolution of some additional BOPs at the birthplace of the crystal nucleus, in \textbf{(a)} $q_2$, \textbf{(b)} $q_7$, \textbf{(c)} $q_8$, \textbf{(d)} $q_9$, and \textbf{(e)} $q_{10}$. 
    The vertical dashed line in each figure indicates the start of nucleation $t_0$, and the shaded area indicates the time window before $t_0$ for which $\text{ACF}(S_2)>0.05$. The left panels, \textbf{(a1-e1)}, show the evolution of the parameters over a long time interval, and the right panels, \textbf{(a2-d2)}, show the same evolutions but now zoomed in on the short time interval around the start of nucleation. In each figure, the black line indicates the number of particles in the crystal nucleus. The horizontal dashed lines indicate the reference value of each parameter in the bulk fluid and crystal phases. The left vertical axis of each figure is scaled in such a way that they all depict the same relative range with respect to these bulk values.
    }
\end{figure*}

Figure \ref{fig:sievent505moreQ} shows, for the event depicted in Fig. 8 of the main paper, the time evolution of some additional BOPs. As for the order parameters discussed in the main paper, we observe that all BOPs depicted in Fig. \ref{fig:sievent505moreQ} increase simultaneously, and in a similar fashion, as soon as nucleation starts. That we can observe the simultaneous increase in these BOPs this clearly is quite astonishing, as the reference values of the bulk crystal and fluid phases for some of these BOPs differ only by $\sim10$\%.

\begin{figure*}[h!]
    \centering
    \begin{tabular}{cc}
        \includegraphics[width=\figwidthB]{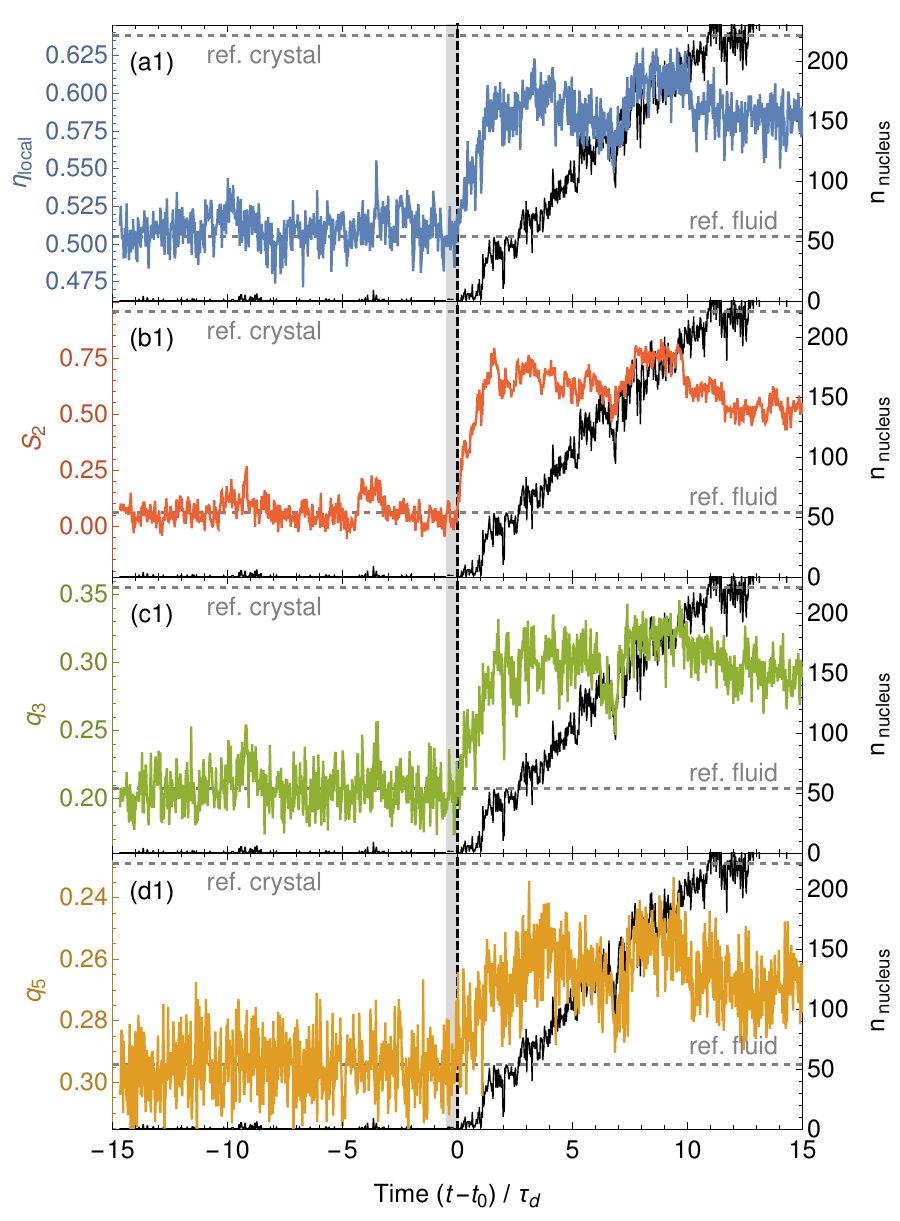}
        & \includegraphics[width=\figwidthZ]{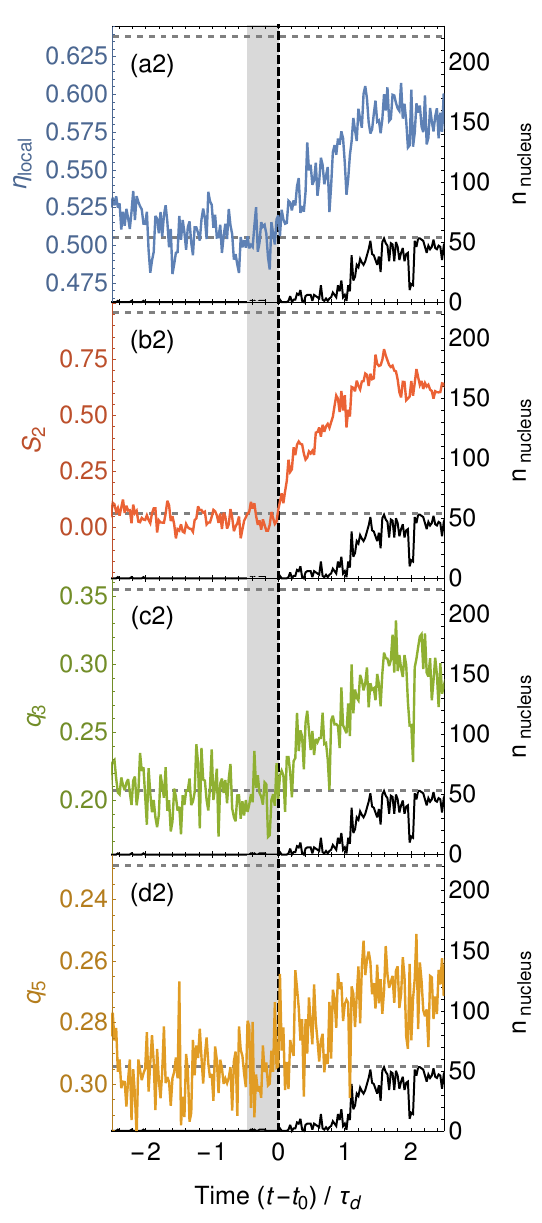}
    \end{tabular}
    \caption[width=1\linewidth]{\label{fig:sievent505} 
    Time evolution of the average local order parameters at the birthplace of the crystal nucleus for another typical nucleation event at packing fraction $\eta=0.505$. 
    The vertical dashed line in each figure indicates the start of nucleation $t_0$, and the shaded area indicates the time window before $t_0$ for which $\text{ACF}(S_2)>0.05$. The left panels, \textbf{(a1-d1)}, show the evolution of the parameters over a long time interval, and the right panels, \textbf{(a2-d2)}, show the same evolutions but now zoomed in on the short time interval around the start of nucleation. In each figure, the black line indicates the number of particles in the crystal nucleus. The colored lines indicate \textbf{(a)} the local packing fraction, \textbf{(b)} the nematic order parameter, and the \textbf{(c)} 3-fold and \textbf{(d)} 5-fold symmetric BOPs. 
    Note that the vertical axis of \textbf{(d)} is inverted such that, even though $q_5$ is decreasing as nucleation progresses, the inverted trend shows a similar increase as for the other order parameters.
    The horizontal dashed lines indicate the reference value of each parameter in the bulk fluid and crystal phases. The left vertical axis of each figure is scaled in such a way that they all depict the same relative range with respect to these bulk values.
    }
\end{figure*}

\begin{figure*}[h]
    \centering
    \begin{tabular}{cc}
        \includegraphics[width=\figwidthB]{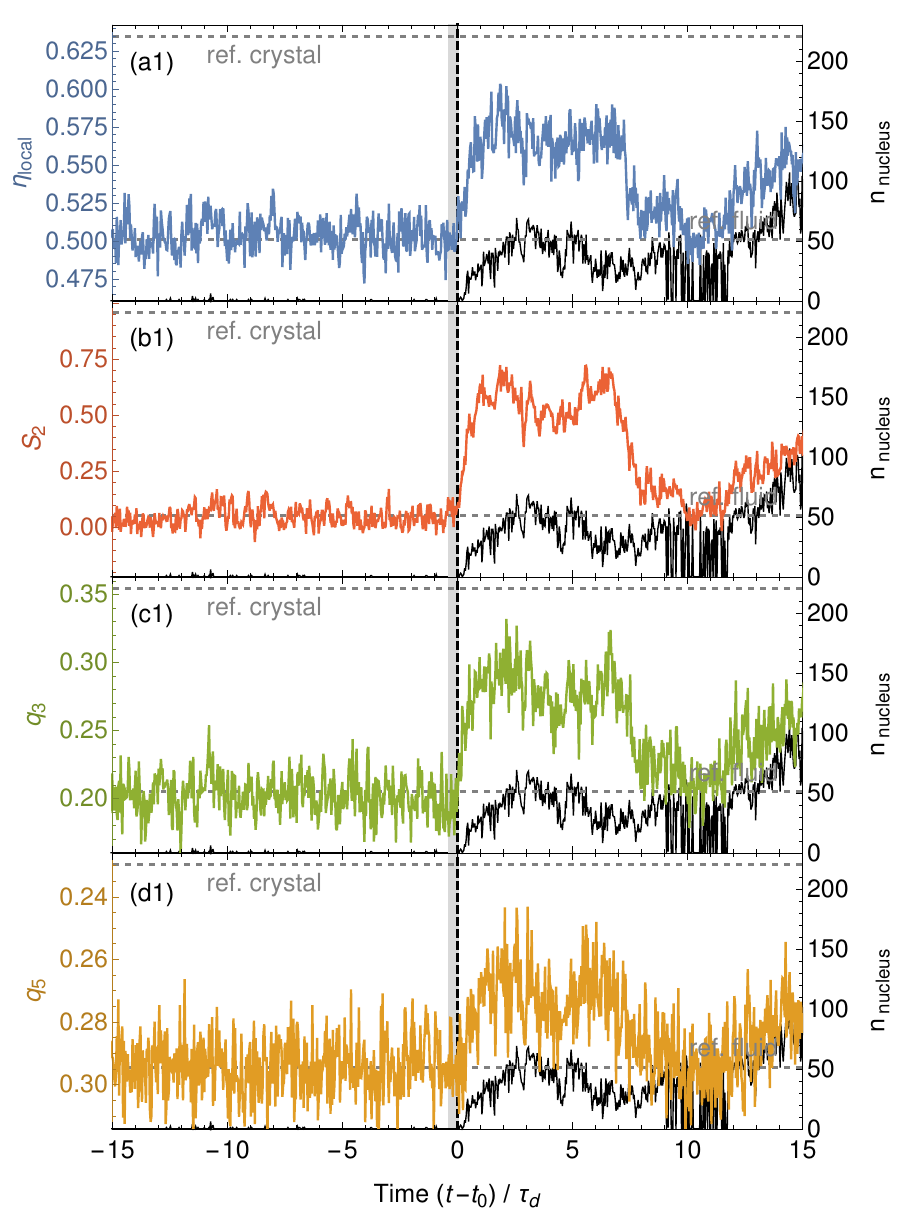}
        & \includegraphics[width=\figwidthZ]{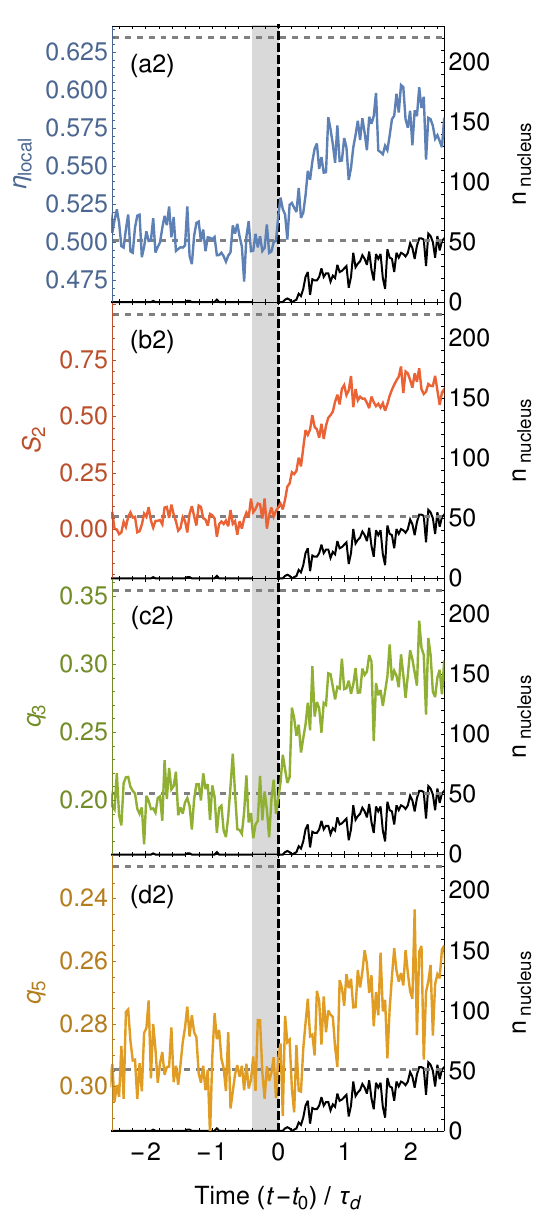}
    \end{tabular}
    \caption[width=1\linewidth]{\label{fig:sievent501} 
    Time evolution of the average local order parameters at the birthplace of the crystal nucleus for another typical nucleation event at packing fraction $\eta=0.501$. 
    See caption of Fig. \ref{fig:sievent505} for an explanation of the figures.
    The initial nucleus melts at $(t-t_0)/\tau_d\sim10$, after which it resumes its growth in a slightly shifted location.
    }
\end{figure*}

\begin{figure*}[h]
    \centering
    \begin{tabular}{cc}
        \includegraphics[width=\figwidthB]{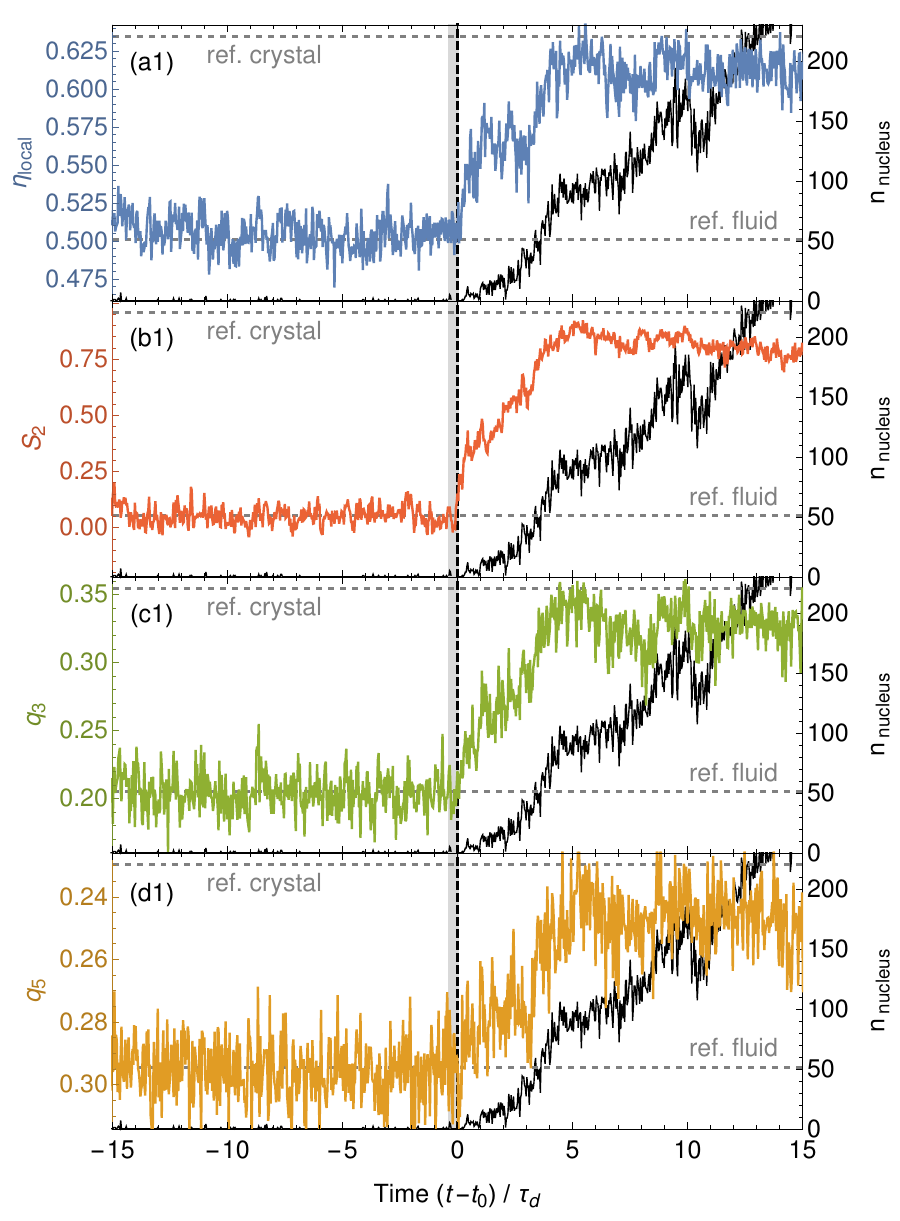}
        & \includegraphics[width=\figwidthZ]{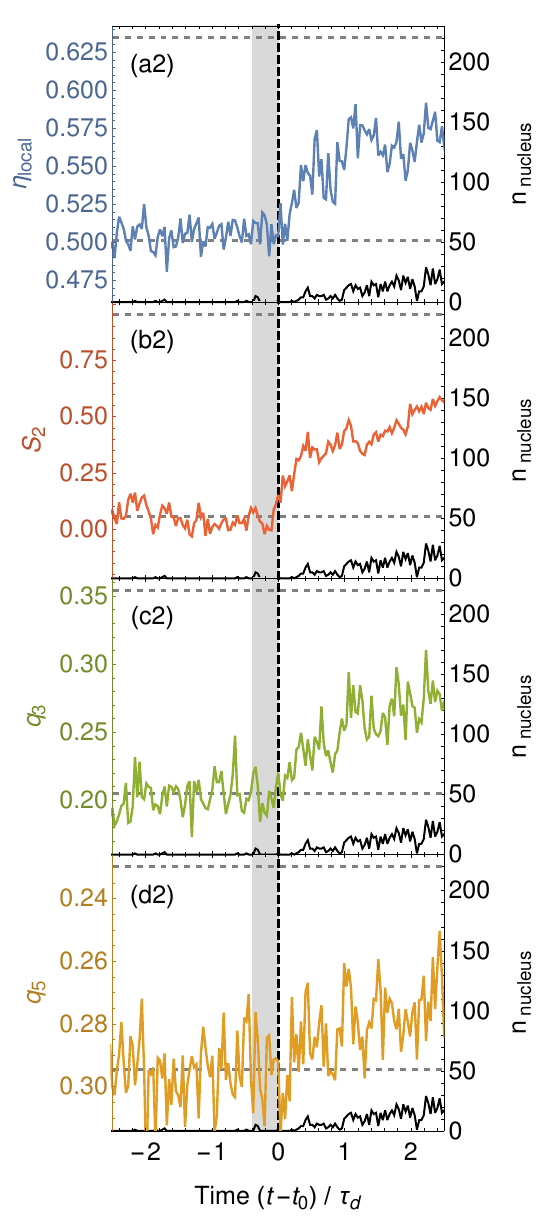}
    \end{tabular}
    \caption[width=1\linewidth]{\label{fig:sievent501v2} 
    Time evolution of the average local order parameters at the birthplace of the crystal nucleus for another typical nucleation event at packing fraction $\eta=0.501$. 
    See caption of Fig. \ref{fig:sievent505} for an explanation of the figures.
    }
\end{figure*}

The last figures of this document show the time evolution of the local order parameters of three additional nucleation events. 
Figure \ref{fig:sievent505} shows an event for $\eta=0.505$ and Figs. \ref{fig:sievent501} and \ref{fig:sievent501v2} show two events for $\eta=0.501$. As for the events in the main text, we observe a simultaneous and abrupt increase in all order parameters. 
Note that, for the event of Fig. \ref{fig:sievent501}, the initial nucleus largely melts before it resumes to grow again. However, as it resumes its growth at a slightly shifted location, the location $\mathbf{r}_0$ of the birthplace of the initial crystal nucleus no longer coincides with the center of mass of the growing nucleus. This causes the drop in the values of the local properties at later times (Fig. \ref{fig:sievent501}a1-d1). Nonetheless, the initial start of nucleation (Fig. \ref{fig:sievent501}a2-d2) behaves the same as for other nucleation events.
Events like the one in Fig. \ref{fig:sievent501} demonstrate why it is impractical to average the time evolution of multiple nucleation events together.

\bibliography{si_paper}

